\documentclass[%
nofootinbib,
 amsmath,amssymb,
 aps,
prm,
twocolumn,
]{revtex4-2}

\usepackage{newtxtext,newtxmath}  
\usepackage{graphicx}
\usepackage{booktabs}
\usepackage{dcolumn} 
\usepackage{bm}
\usepackage{microtype}
\usepackage{xcolor}
\definecolor{darkblue}{rgb}{0,0,0.6}
\definecolor{apsblue}{rgb}{0.18,0.19,0.57}
\usepackage[colorlinks,linkcolor=apsblue,citecolor=apsblue,urlcolor=apsblue]{hyperref}

\newcommand{\R}[2]{R[{#1}, {#2}]}
\newcommand{\Rb}[2]{{\bm R}[{#1}, {#2}]}
\newcommand{\mean}[1]{{\mathbb{E}}[#1]}

\graphicspath{{figures/}:{./}}

\begin{document}

\title{
  Interpretability of linear regression models of glassy dynamics
}

\author{Anand Sharma}
\affiliation{Indian Institute of Science Education and Research,
Dr. Homi Bhabha Road, Pashan, Pune 411008, India}

\affiliation{Univ. Grenoble Alpes, CNRS, LIPhy, 38000 Grenoble, France}

\affiliation{Department of Physics and Materials Science, University of Luxembourg, L-1511 Luxembourg, Luxembourg}

\author{Chen Liu}
\affiliation{Innovation and Research Division, Ge-Room Inc., 93160 Noisy le Grand, France}

\author{Misaki Ozawa}
\affiliation{Univ. Grenoble Alpes, CNRS, LIPhy, 38000 Grenoble, France}

\email{* misaki.ozawa@univ-grenoble-alpes.fr}
\author{Daniele Coslovich}
\email[Corresponding author: ]{dcoslovich@units.it}
\affiliation{Dipartimento di Fisica, Universit\`a di Trieste, Strada Costiera 11, 34151, Trieste, Italy}

\date{\today}

\begin{abstract}

  Data-driven models can accurately describe and predict the dynamical properties of glass-forming liquids from structural data. Accurate predictions, however, do not guarantee an understanding of the underlying physical phenomena and the key factors that control them. In this paper, we illustrate the merits and limitations of linear regression models of glassy dynamics built on high-dimensional structural descriptors. By analyzing data for a two-dimensional glass model, we show that several descriptors commonly used in glass‑transition studies display multicollinearity,
  which hinders the interpretability of linear models. Ridge regression suppresses some of the shortcomings of multicollinearity, but its solutions are not concise enough to be physically interpretable. Only by using dimensional reduction techniques we do eventually obtain linear models that strike a balance
 between prediction accuracy and interpretability.
Our analysis points to a key role of local packing and composition fluctuations in the glass model under study.

\end{abstract}

\maketitle

\section{Introduction}

As the temperature of glass-forming liquids decreases, the relaxation timescale increases dramatically, while structural changes remain relatively modest~\cite{berthier2011theoretical,binder2011glassy,ediger1996supercooled}. Moreover, the dynamics exhibits pronounced spatial fluctuations, characterized by regions of high and low mobility, referred to as dynamic heterogeneities~\cite{ediger2000spatially,berthier2011dynamical,karmakar2014growing}. Despite the vivid patterns of dynamic heterogeneities, however, static snapshots of the system appear homogeneous and lack distinct features, at least to the naked eye. This has been confirmed through direct observations in computer simulations and colloidal glass experiments~\cite{widmer2004reproducible,weeks2017introduction}.
To identify subtle but significant structural changes linked to dynamics, various structural order parameters based on physical intuition have been investigated~\cite{coslovich2007understanding, patrick2008direct, widmer2008irreversible,  hocky2014correlation, jack2014information, turci2017nonequilibrium, royall2015role, tanaka2019revealing,kapteijns2021does,sharma2022identifying}.
Similar efforts have been made to identify the structural origins of plastic events in amorphous solids under loading~\cite{manning2011vibrational, richard2020predicting, baggioli2021plasticity, wu2023topology}.

Recent advances in machine learning have demonstrated that glassy dynamics, including dynamic heterogeneities, can be accurately described and predicted from local structural information~\cite{cubuk2015identifying,jung2025roadmap}. A range of machine learning techniques has been applied to this problem, including support vector machines~\cite{cubuk2015identifying,schoenholz2016structural}, multi-layer perceptrons~\cite{jung2023predicting}, and graph neural networks~\cite{bapst2020unveiling,shiba2023botan,jiang2023geometry,pezzicoli2024rotation}. This research has primarily progressed in two directions. The first involves increasing the complexity of machine learning architectures, employing deep neural networks with a large number of parameters and taking advantage of cutting-edge methods~\cite{bapst2020unveiling,shiba2023botan,pezzicoli2024rotation}. The second direction focuses on integrating domain knowledge from physics in data-driven models~\cite{boattini2021averaging,jung2023predicting}. These physics-informed approaches enable high predictive accuracy while maintaining relatively simple model architectures.

However, achieving accurate predictions alone does not guarantee an understanding of the underlying mechanisms driving the phenomenon under investigation~\cite{teney2022predicting,swain2024machine}. From the perspective of fundamental physics research, it is crucial that data-driven models provide physically interpretable results, which must be robust and expressed in a succinct form.
A growing body of glass transition studies explores the issue of interpretability in deep‑learning models or non-linear models, using a range of approaches~\cite{swanson2020deep,font2022predicting,oyama2023deep,ciarella2023finding,janzen2023dead,janzen2024classifying,liu2024classification,jung2024dynamic,wang2024predicting,yoshikawa2025graph,wang2020predicting,liu2023concurrent}.
While these works offer useful clues, the underlying deep networks remain highly complex and only partially transparent.
Consequently, there is still a clear need for explicitly interpretable models whose solutions admit a direct physical reading.

Remarkably, recent studies have demonstrated that simple linear models, when combined with domain knowledge of glassy dynamics, such as coarse-graining techniques, can describe dynamic heterogeneities with remarkable accuracy~\cite{boattini2021averaging,alkemade2022comparing}. In some cases, their accuracy is comparable to that of more complex deep learning models. This outcome is particularly desirable, because it is often believed that simpler models not only provide accurate predictions at a minimal computational cost, but can also offer greater interpretability~\cite{rudin2019stop}.
We caution, however, that employing a linear model {\it per se} is not sufficient to extract meaningful physical information: in high‑dimensional settings, linear models are often plagued by numerical instabilities, because of so-called multicollinearity~\cite{montgomery2021introduction},
that can obscure interpretation.
Moreover, their solution may still be too high-dimensional to convey a physical meaning.
The aim of this study is to address these issues in linear models for glassy dynamics and introduce remedies that restore interpretability.

To tackle the issue of interpretability in a rigorous manner, 
we will quantitatively study the consequences of multicollinearity in linear regression models of glassy dynamics, using a two-dimensional glass-forming liquid model.
First, we will demonstrate how multicollinearity leads to instability in weight estimation within linear regression, thereby hindering the interpretation of feature importance.
These problems affect several structural descriptors used in recent glass transition studies.
Next, we will explore strategies to mitigate the effects of multicollinearity and to identify a set of low-dimensional linear models that achieve good performance accuracy with a minimum of structural information.
We will critically evaluate and discuss the advantages and limitations of each approach, providing a comprehensive assessment of linear models of the structure-dynamics relationship in a glass-forming liquid model.

The paper is organized as follows.
Section~\ref{sec:problem} defines the problem under investigation.
Section~\ref{sec:simulation} describes the physical model and our structure-dynamics dataset.
Sections~\ref{sec:linear_regression} and~\ref{sec:multicollinearity} introduce simple linear regression models of glassy dynamics and demonstrate the effects of multicollinearity on the instability of the estimated weights.
Section~\ref{sec:resolutions} examines feature selection and extraction approaches that cope with multicollinearity while also reducing the dimensionality of the problem.
Sections~\ref{sec:discussion} and~\ref{sec:conclusions} offer a critical outlook on our results and summarize the key findings of the work.

\section{Problem statement}
\label{sec:problem}

In data-driven modeling of glassy dynamics, the problem is typically formulated as follows~\cite{jung2025roadmap}. Several variables, called features, are computed in order to characterize the local structural environment around each particle.
These structural features are collected into a descriptor that provides a high-dimensional representation of the local structure.
Linear regression models estimate the dynamical observable ${\bf Y}$ (in this study, the dynamic propensity~\cite{widmer2006predicting}) as a linear combination of $M$ input structural features, ${\bf X}^{(1)}, {\bf X}^{(2)}, \ldots, {\bf X}^{(M)}$, using the following expression:
\begin{equation}
\hat {\bf Y} = \hat w^{(1)} {\bf X}^{(1)} + \hat w^{(2)} {\bf X}^{(2)} + \cdots + \hat w^{(M)} {\bf X}^{(M)},
\label{eq:linear_model_intro}
\end{equation}
where $\hat w^{(1)}, \hat w^{(2)}, \ldots, \hat w^{(M)}$ are the weights determined by minimizing a specific loss function.

How can we extract physical insights from the linear model in Eq.~\eqref{eq:linear_model_intro}?
When ${\bf Y}$ and ${\bf X}^{(f)}$ ($f=1, 2, \ldots, M$) are properly normalized, for example, to have zero mean and unit variance, the sign and magnitude of the weights $\hat w^{(f)}$ provide information about the influence of the corresponding input feature ${\bf X}^{(f)}$ on the dynamical output ${\bf Y}$ (and its prediction $\hat {\bf Y}$). In other words, $\hat w^{(f)}$ serves as a measure of feature importance.
Thus, linear regression models provide a strong case for physical interpretability, in which the connection between ${\bf X}$ and ${\bf Y}$ via the weights is direct and mechanistic.

In practice, however, the estimation of $\hat w^{(f)}$ can be unstable if ${\bf X}^{(1)}, {\bf X}^{(2)}, \ldots, {\bf X}^{(M)}$ are strongly correlated with one another.
As an extreme case, if ${\bf X}^{(1)}, {\bf X}^{(2)}, \ldots, {\bf X}^{(M)}$ are indeed linearly dependent, the weights $\hat w^{(1)}, \hat w^{(2)}, \ldots, \hat w^{(M)}$ are not uniquely determined.
In real datasets, situations close to linear dependency frequently arise.
Then, small perturbations in the dataset, such as numerical errors or limited statistics, can cause large variations in the weights. This pathological yet common phenomenon is known as multicollinearity~\cite{montgomery2021introduction}.

Typically, multicollinearity is not a major concern in machine-learning studies, where the primary objective is to optimize predictive accuracy. It becomes problematic, however, when regression models are interpreted in terms of feature importance.
In particular, if the estimation of the regression weights is unstable, any physical interpretation becomes unreliable.
Two distinct mechanisms can undermine the stability of the estimated weights because of multicollinearity.
The first one corresponds to large dataset-to-dataset fluctuations: the values of the weights may vary significantly depending on the particular subsample of data used for training, even when all datasets are generated under identical physical conditions.
In the large-sample regime, this variance is suppressed, leading to robust and reproducible weight estimates. A second, qualitatively different pathological behavior persists even in the large-sample limit.
In this case, the weights exhibit oscillatory behavior across features.
This instability originates also from multicollinearity.
Note that this effect is not associated with statistical uncertainty but with the geometry of the feature space itself, and therefore cannot be mitigated by increasing the dataset size. Robustness of the estimated weights against both sources of instability is a basic requirement for interpretability of data-driven models.

The very notion of interpretability is context dependent~\cite{broniatowski2021psychological} and remains the subject of active debate; see Ref.~\onlinecite{Wetzel_Ha_Iten_Klopotek_Liu_2025} for a recent review in the physics context.
Recent works have emphasized the role of sparsity~\cite{Rowan_Doostan_2025}, whereby the relevant features of a data-driven model stand out clearly from the bulk, \textit{e.g.}, in terms of their weights or mutual distances in feature space.
In our view, physically interpretable models must additionally provide a concise and robust description of the relationships between the variables of interest.
This perspective is consistent with the physicists' expectation that successful models strongly compress the information contained in the problem~\cite{kivelson2018understanding}.
Phenomenological models in liquid-state theory and statistical physics, for instance, are often built from a small number of independent variables, such as in hydrodynamic descriptions or two-state models.
Accordingly, to be physically interpretable, data-driven models must identify robust relationships involving a limited set of relevant variables, each carrying a clear physical meaning.

To close this section, a comment on terminology is in order.
To evaluate the quality of a data-driven model, one computes the estimate $\hat {\bf Y}$ for a set of input data that were not used to train the model.
In the following, we will do that by splitting the original data set into a training and test set, as is customary.
The estimate $\hat {\bf Y}$ is then considered as prediction for the test set; standard goodness-of-fit metrics can be used to evaluate their quality.
From a conventional physicist's viewpoint, this is a weak form of prediction: in the problem we will be dealing with, the data points will be sampled at precisely the same physical conditions both in the train and test set and no attempt will be made to extrapolate to different conditions. 
We will nonetheless stick to the word ``prediction'' throughout this work, for consistency with previous work~\cite{jung2025roadmap} and with the widespread usage in the machine learning context.
Should the readers feel uncomfortable with it, they can mentally replace ``prediction'' with ``description'' in the following.
Only in Sec.~\ref{sec:discussion} we will assess the cross-state predictions of the models, by extrapolating the dynamic propensity to different temperatures and time scales.

\section{Simulation model and dataset}
\label{sec:simulation}

\subsection{Simulation model}

We use a three-component glass-forming liquid model composed of small (S), medium (M), and large (L) particles in two spatial dimensions with periodic boundary conditions~\cite{sharma2024selecting}.
The model is a variant of a well-studied binary mixture model~\cite{falk1998dynamics,barbot2018local,barbot2018local,lerbinger2022relevance}, to which we add particles of species M with an intermediate character between S and L~\cite{parmar2020ultrastable}.
The interaction between two particles is described by the Lennard-Jones potential
\begin{equation}
v_{\alpha \beta}(r) = 4 \epsilon_{\alpha \beta} \left[\left(\frac{\sigma_{\alpha \beta}}{r}\right)^{12} - \left(\frac{\sigma_{\alpha \beta}}{r}\right)^6 \right],
\end{equation}
where $\alpha, \beta = \mathrm{S}, \mathrm{M}, \mathrm{L}$. The potential is modified to ensure that it is twice continuously differentiable at the cutoff, following Ref.~\onlinecite{barbot2018local}.

The parameters $\sigma_{\alpha \beta}$ and $\epsilon_{\alpha \beta}$ are:  
$$
\begin{aligned}
& \sigma_{\mathrm{LL}} = 2 \sin\left(\frac{\pi}{5}\right) \simeq 1.18, \ \sigma_{\mathrm{SS}} = 2 \sin\left(\frac{\pi}{10}\right) \simeq 0.62, \ \sigma_{\mathrm{LS}} = 1, \\
& \sigma_{\mathrm{LM}} = \frac{\sigma_{\mathrm{LL}} + \sigma_{\mathrm{LS}}}{2}, \ \sigma_{\mathrm{MS}} = \frac{\sigma_{\mathrm{LS}} + \sigma_{\mathrm{SS}}}{2}, \ \sigma_{\mathrm{MM}} = \frac{\sigma_{\mathrm{LL}} + \sigma_{\mathrm{SS}}}{2}, \\
& \epsilon_{\mathrm{LL}} = \frac{1}{2}, \ \epsilon_{\mathrm{SS}} = \frac{1}{2}, \ \epsilon_{\mathrm{LS}} = 1, \\
& \epsilon_{\mathrm{LM}} = \frac{\epsilon_{\mathrm{LL}} + \epsilon_{\mathrm{LS}}}{2}, \ \epsilon_{\mathrm{MS}} = \frac{\epsilon_{\mathrm{LS}} + \epsilon_{\mathrm{SS}}}{2}, \ \epsilon_{\mathrm{MM}} = \frac{\epsilon_{\mathrm{LL}} + \epsilon_{\mathrm{SS}}}{2}.
\end{aligned}
$$  
The total number of particles is $N = N_{\mathrm{S}} + N_{\mathrm{M}} + N_{\mathrm{L}} = 4000$, where $N_{\mathrm{S}} = 1760$, $N_{\mathrm{M}} = 800$, and $N_{\mathrm{L}} = 1440$ are the numbers of small, medium, and large particles, respectively.  
We use the $NVT$ canonical ensemble, with a number density $\rho = N / L^2 = 1.024$, where $L$ is the linear length of the square simulation cell.  

We perform Monte Carlo (MC) simulations using translational displacements~\cite{frenkel2023understanding}.
The MC move consists in picking a particle at random and displacing it by a vector drawn randomly within a square box of linear size $\delta_{\rm max} = 0.12$.
The move is accepted on the basis of the Metropolis acceptance rule, which ensures the detailed balance condition.
Although MC simulations do not possess a physical timescale, time $t$ can be measured in units of MC sweeps, each comprising $N$ attempts to perform the MC move.
In the regime of slow glassy dynamics of interest in this work, the Monte Carlo dynamics behaves similarly to other types of physical dynamics, \textit{e.g.}, Newtonian and Brownian dynamics~\cite{berthier2007monte}.
Therefore, we analyze the MC dynamics by following particle trajectories and calculating time-dependent observables as usual. 
The glassy dynamics of the model has been studied in Ref.~\onlinecite{sharma2024selecting} by computing the self intermediate scattering function. 
In this study, we mainly focus on $T=0.30$, which is the lowest temperature at which we can equilibrate the system within our computational timescale.

\subsection{Dynamic propensity}

To investigate the heterogeneity of glassy dynamics in real space and assess its connection with the static structure, we use the isoconfigurational ensemble~\cite{widmer2006predicting,widmer2008irreversible}.
A set of $n=10$ statistically uncorrelated configurations are obtained at equilibrium conditions at temperature $T = 0.30$.
From each of these equilibrium configurations, we generate an ensemble of trajectories using MC dynamics at $T = 0.30$ using 60 different initial random seeds. 
We then compute the dynamic propensity
\begin{equation}
p_i(t) = \left\langle |\Delta {\bf r}^{\rm CR}_i(t)| \right\rangle_{\rm iso} , 
\end{equation}
where $\langle \cdots \rangle_{\rm iso}$ denotes an average over all the trajectories originating from the same initial configuration, and
the cage-relative displacement, $\Delta {\bf r}^{\rm CR}_i(t)$, is given by
$$
\Delta {\bf r}^{\rm CR}_i(t) = \Delta {\bf r}_i(t) - \frac{1}{n_i} \sum_{j \in \mathcal{N}_i} \Delta {\bf r}_j(t),
$$  
where $\Delta {\bf r}_i(t) = {\bf r}_i(t) - {\bf r}_i(0)$ is the displacement vector of the $i$-th particle at position ${\bf r}_i$.
Here, $n_i$ is the number of neighboring particles, and the set of neighbors ($\mathcal{N}_i$) is defined as the particles located within a circular cutoff radius of $1.4 \sigma_{\alpha \beta}$.  
The choice of cage-relative displacements is necessary to filter out the effect of the so-called Mermin-Wagner fluctuations in two-dimensional systems~\cite{shiba2016unveiling}.
In this paper, we focus on the dynamic propensity computed at the structural relaxation timescale, $p_i(\tau_\alpha)$, where $\tau_\alpha$ is defined as the time at which the self intermediate scattering function becomes $1/e$~\cite{sharma2024selecting}.
$\tau_\alpha$ at $T=0.30$ is $\tau_\alpha \simeq 4 \times 10^6$.
We also considered a shorter time scale, $t=5 \times 10^4$, which corresponds to $\beta$ relaxation timescale.
The main findings of this work remained qualitatively unchanged.

\subsection{Behler-Parrinello descriptor}
\label{sec:behler_parrinello}

To characterize the local structure around each particle, we use the Behler-Parrinello (BP) descriptor~\cite{behler2007generalized}, which has been widely used to study structure-property relationships, including the description of glassy dynamics in two dimensions~\cite{cubuk2015identifying, rocks2021learning}.
The descriptor comprises two subsets of features that characterize radial and angular correlations, respectively.
Following Refs.~\onlinecite{cubuk2015identifying, rocks2021learning}, we further distinguish features according to the species of the neighboring particles.

For each particle $i$, the radial feature $G_i^\alpha$ is defined by
\begin{equation}
    G_i^\alpha = \sideset{}{'}\sum_{j \in \mathcal{N}_\alpha} e^{-(r_{ij}-\mu)^2/\delta^2} f_c(r_{ij}) ,
\end{equation}
where $r_{ij}=|{\bf r}_i-{\bf r}_j|$ is the distance between particles $i$ and $j$, and  $\mu$ and $\delta$ are parameters.
The sum is carried out over the subset $\mathcal{N}_\alpha$ of particles of species $\alpha$. 
$\sum'$ indicates that the particle $i$ is removed from the sum.
The cut-off function $f_c(r)$ is defined by $f_c(r)=\frac{1}{2}\left[ \cos(\pi r/R_c) + 1\right]$ for $r \leq R_c$ and $f_c(r)=0$ for $r>R_c$~\cite{behler2015constructing}. $R_c$ is a cut-off radius and we set $R_c=5.0 \sigma_{\rm LS}$.
We vary $\mu$ between $0.3 \sigma_{\rm LS}$ and $5.0 \sigma_{\rm LS}$ in increments of $0.1\sigma_{\rm LS}$ with $\delta=0.1\sigma_{\rm LS}$.
Thus, for each species $\alpha$, the radial features $G^\alpha(k)$ are parametrized by an integer $k$ that selects values of the parameter $\mu = \mu_k = 0.3\sigma_{\rm LS} + k \times 0.1\sigma_{\rm LS}$ and $0\le k \le 47$.
Thus, for each particle, we have $144(=3 \times 48)$ different radial features.

\begin{table}
  \centering
  {\renewcommand{\arraystretch}{1.1} \begin{tabular}{lrrr}
\hline
\hline
&$\xi$&$\zeta$&$\lambda$\\[0pt]
\hline
$\Psi^{\alpha\beta}(0)$&$14.633$&$1$&$-1$\\[0pt]
$\Psi^{\alpha\beta}(1)$&$14.633$&$1$&$1$\\[0pt]
$\Psi^{\alpha\beta}(2)$&$14.638$&$2$&$-1$\\[0pt]
$\Psi^{\alpha\beta}(3)$&$14.638$&$2$&$1$\\[0pt]
$\Psi^{\alpha\beta}(4)$&$2.554$&$1$&$-1$\\[0pt]
$\Psi^{\alpha\beta}(5)$&$2.554$&$1$&$1$\\[0pt]
$\Psi^{\alpha\beta}(6)$&$2.554$&$2$&$-1$\\[0pt]
$\Psi^{\alpha\beta}(7)$&$2.554$&$2$&$1$\\[0pt]
$\Psi^{\alpha\beta}(8)$&$1.648$&$1$&$1$\\[0pt]
$\Psi^{\alpha\beta}(9)$&$1.648$&$2$&$1$\\[0pt]
$\Psi^{\alpha\beta}(10)$&$1.204$&$1$&$1$\\[0pt]
$\Psi^{\alpha\beta}(11)$&$1.204$&$2$&$1$\\[0pt]
$\Psi^{\alpha\beta}(12)$&$1.204$&$4$&$1$\\[0pt]
$\Psi^{\alpha\beta}(13)$&$1.204$&$16$&$1$\\[0pt]
$\Psi^{\alpha\beta}(14)$&$0.933$&$1$&$1$\\[0pt]
$\Psi^{\alpha\beta}(15)$&$0.933$&$2$&$1$\\[0pt]
$\Psi^{\alpha\beta}(16)$&$0.933$&$4$&$1$\\[0pt]
$\Psi^{\alpha\beta}(17)$&$0.933$&$16$&$1$\\[0pt]
$\Psi^{\alpha\beta}(18)$&$0.695$&$1$&$1$\\[0pt]
$\Psi^{\alpha\beta}(19)$&$0.695$&$2$&$1$\\[0pt]
$\Psi^{\alpha\beta}(20)$&$0.695$&$4$&$1$\\[0pt]
$\Psi^{\alpha\beta}(21)$&$0.695$&$16$&$1$\\[0pt]
\hline
\hline
\end{tabular}}

  \caption{\label{table:features_bp_angular}Parameters of the angular features $\Psi^{\alpha\beta}(k)$ of the BP descriptor.
  }
\end{table}

The angular descriptor $\Psi_i^{\alpha\beta}$ is defined by
\begin{eqnarray}
\Psi_i^{\alpha\beta} = 2^{1-\zeta} \sideset{}{'}\sum_{\substack{{j \in \alpha}, \ {k \in \beta}\\ (j \neq k)}} e^{-(r_{ij}^2+r_{ik}^2+r_{jk}^2)/\xi^2} \nonumber \\ \times (1+\lambda \cos \theta_{ijk})^\zeta  f_c(r_{ij}) f_c(r_{ik}) f_c(r_{jk}) ,
\end{eqnarray}
where $\theta_{ijk}$ is the angle at the corner $i$ of the triangle defined by particles $i$, $j$, and $k$, and $\xi$, $\lambda$, and $\zeta$ are parameters that are varied systematically.
For each pair of species, $(\alpha, \beta)$, we employ the same set of $22$ parameters ($\xi$, $\lambda$, $\zeta$) given in Ref.~\onlinecite{cubuk2015identifying} in unit of $\sigma_{\rm LS}$.
The features $\Psi^{\alpha\beta}(k)$ are parametrized by an integer $k$ and the parameters $(\xi=\xi_k, \lambda=\lambda_k, \zeta=\zeta_k)$ with $0\le k \le 21$ are shown in Table~\ref{table:features_bp_angular}.
Thus, we have $132 (=6 \times 22)$ angular features.

The full BP descriptor comprises a total of $M=276(=144+132)$ features for each particle.
Contrary to previous work~\cite{cubuk2015identifying, rocks2021learning}, we coarse-grain each feature over a length scale $\ell=1.5$ using the procedure described in Sec.~\ref{sec:physical_descriptors}. Coarse-graining improves the prediction accuracy of the descriptor, without changing qualitatively the conclusions of this work.
These variables constitute a feature vector, given by
\begin{equation}
    {\bf X}_i=\left(X_i^{(1)}, X_i^{(2)}, \cdots, X_i^{(M)}\right) .
\end{equation}
We will sort the different kinds of features as follows:
\begin{equation}
  {\bf X}_i=\left( G_i^{\rm S}, G_i^{\rm M}, G_i^{\rm L}, \Psi_i^{\rm SS}, \Psi_i^{\rm SM}, \Psi_i^{\rm SL}, \Psi_i^{\rm MM}, \Psi_i^{\rm ML}, \Psi_i^{\rm LL} \right) .
\end{equation}

\subsection{Physically motivated descriptors}
\label{sec:physical_descriptors}

To assess the generality of our findings, we also consider two additional structural descriptors that have been recently used to study structure-dynamics relationships in glass-forming liquids~\cite{sharma2024selecting, jung2023predicting}.
Both of them are physically motivated: they are based on single-particle structural variables that characterize the environment around a particle in a physically intuitive way.
All these single-particle variables are coarse-grained~\cite{boattini2021averaging} over multiple length scales, as described at the end of this section, to compose the full descriptor.

The local potential energy $u_i$ for particle $i$ is defined by
\begin{equation}
u_i = \frac{1}{2} \sum_{j \neq i} v_{\alpha_i \beta_j}(r_{ij}),
\end{equation}
where $v_{\alpha_i \beta_j}(r_{ij})$ is the pair-wise Lennard-Jones potential, with a cutoff at $2.5 \sigma_{\alpha_i \beta_j}$.

The coordination number $z_i$ for particle $i$ is defined as the number of neighboring particles within $r_{ij} < 1.5 \sigma_{\alpha_i \beta_j}$, which corresponds well to the first minimum of each partial radial distribution function, $g_{\alpha\beta}(r)$.

The bond-orientational order parameter in two dimensions, $\Psi_{6, i}$, is defined by
\begin{equation}
\Psi_{6, i} = \frac{1}{z_i} \left| \sum_{j=1}^{z_i} e^{\sqrt{-1} \ 6 \theta_{ij}} \right|,
\end{equation}
where $\theta_{ij}$ is the angle between ${\bf r}_{ij} = {\bf r}_j - {\bf r}_i$ and the $x$ axis. The nearest neighbors are again defined as those within $r_{ij} < 1.5 \sigma_{\alpha_i \beta_j}$. $\Psi_{6, i}$ quantifies hexagonal order, taking the value $1$ for perfect hexagonal packings and smaller values for disordered packings~\cite{kawasaki2007correlation,schreck2011tuning}.

The steric bond order parameter $\Theta_i$~\cite{tong2018revealing} is a measure of how well packed is the local environment around particle $i$.
For each pair $\langle jk \rangle$ of neighboring particles, the angle $\theta_{jk}$ between ${\bf r}_{ij}$ and ${\bf r}_{ik}$ is compared to the reference angle $\theta_{jk}^{\rm ref}$, calculated using the cosine formula. The steric order parameter is given by
\begin{equation}
\Theta_i = \frac{1}{z_i} \sum_{\langle jk \rangle} \left| \theta_{jk} - \theta_{jk}^{\rm ref} \right|,
\end{equation}
where $\langle jk \rangle$ denotes the summation over all pairs of neighbors. Smaller values of $\Theta_i$ indicate sterically favored configurations, while larger values reflect disordered packings.

The local number density is given by
\begin{equation}
\overline \rho_i(\ell) = \sum_{j \in \mathcal{N}_i} e^{-r_{ij}/\ell},
\end{equation}
where $\mathcal{N}_i$ includes particle $i$, $\ell$ is a coarse-graining length and all the other $N_i$ particles are included in the sum.

The local volume fraction is defined by
\begin{equation}
\overline \varphi_i(\ell) = \frac{1}{\overline \rho_i(\ell)} \sum_{j \in \mathcal{N}_i} (\sigma_{\alpha_i \beta_j})^2 e^{-r_{ij}/\ell}.
\end{equation}

Finally, we also consider the perimeter $\pi_i$ of the Voronoi cell surrounding particle $i$, as obtained from a radical Voronoi tessellation~\cite{rycroft2009voro++}, using the nominal interaction parameters $\sigma_{\alpha\alpha}$ as particle radii.

With these structural features at hand, we can proceed to define two physically motivated descriptors.
First, we define the SLO descriptor from the paper by Sharma, Liu, and Ozawa~\cite{sharma2024selecting}. The SLO descriptor comprises $\overline \rho_i$, $\overline \varphi_i$, $u_i$, $z_i$, $\Psi_{6, i}$, and $\Theta_i$.
Each structural feature, $x_i = u_i, z_i, \Psi_{6, i}, \Theta_i$, is coarse-grained using the procedure
\begin{equation}
\overline x_i(\ell) = \frac{1}{\overline \rho_i(\ell)} \sum_{j \in \mathcal{N}_i} x_j e^{-r_{ij}/\ell} ,
\label{eq:CG1}    
\end{equation}
yielding $\overline u_i(\ell)$, $\overline z_i(\ell)$, $\overline \Psi_{6, i}(\ell)$, and $\overline \Theta_i(\ell)$.
The coarse-graining length scale $\ell$ is varied from $0.5\sigma_\textrm{LS}$ to $5.0\sigma_\textrm{LS}$.
Each kind of physically motivated feature $X(k)$ is parametrized by an integer $k$, yielding features with $\ell = \ell(k) = 0.5\sigma_\textrm{LS} + k \times 0.5\sigma_\textrm{LS}$ and $0\le k \le 9$.
The same lengths are used for the calculation of $\overline \rho_i(\ell)$ and $\overline  \varphi_i(\ell)$.
This yields a total of $M=60$ features.

We also use the JBB descriptor introduced in the paper by Jung, Biroli, and Berthier~\cite{jung2023predicting}.
This descriptor is based on $\overline \rho_i$, $u_i$, and $\pi_i$.
In this work, we ignore the variance of the potential energy, which was included in Ref.~\onlinecite{jung2023predicting}.
The JBB descriptor incorporates particle‑species information more explicitly, as shown below.
A first set of coarse-grained features is obtained using the same procedure as for the SLO descriptor, considering the whole set of neighbors, irrespective of their species, with 10 different coarse-graining length scales $\ell$.
In addition, the JBB descriptor includes coarse-grained features obtained by a procedure similar to Eq.~\eqref{eq:CG1} but taking species into account:
\begin{equation}
\overline x_i^\alpha(\ell) = \frac{1}{\overline \rho_i^{\alpha} (\ell)} \sum_{j \in \mathcal{N}_i^{\alpha}} x_j e^{-r_{ij}/\ell} ,
\label{eq:CG2}    
\end{equation}
where
\begin{equation}
\overline \rho_i^\alpha (\ell) = \sum_{j \in \mathcal{N}_i^\alpha} e^{-r_{ij}/\ell},
\end{equation}
and the sums are restricted to neighbors of species $\alpha$.
Thus, in addition to the species-independent coarse-graining defined by Eq.~\eqref{eq:CG1}, we consider coarse-graining based on Eq.~\eqref{eq:CG2} using the three species of particles ($\alpha=$ S, M, and L).
This yields a total of $M=120$ features (3 descriptors $\times$ 4 types $\times$ 10 length scales).

\begin{figure*}[!htp]
\includegraphics[width=\linewidth]{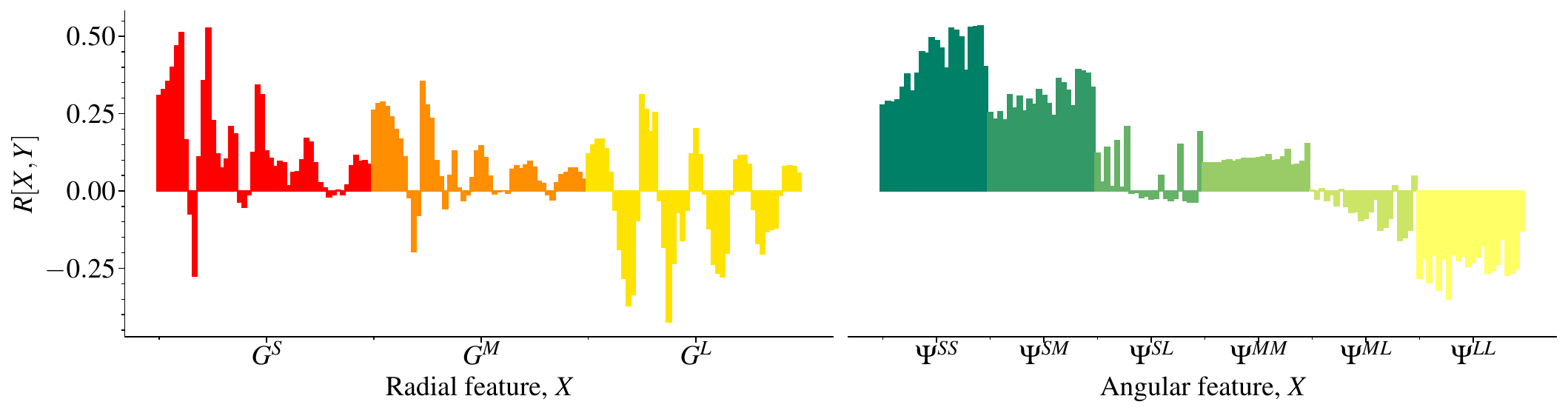}  
\caption{ 
Pearson coefficient, $\R{X^{(f)}}{Y}$, between the dynamic propensity $\textbf{Y}$ and each structural feature $\textbf{X}^{(f)}$ of the BP descriptor for $f=1, \dots, M$.
}
\label{fig:pearson}
\end{figure*}

\subsection{Dataset}
\label{sec:dataset}

The dataset we use in this work comprises both dynamic and structural information.
The propensity is computed using all the $n$ available configurations.
The structural descriptors are instead computed on the inherent structures~\cite{stillinger1984packing} of the initial configurations used for the propensity calculations.
The dataset is then composed of one of the structural descriptors defined in Secs.~\ref{sec:behler_parrinello} and \ref{sec:physical_descriptors} and the dynamic propensity.

As commonly done in machine learning studies, the dataset is feature scaled.
In particular, we normalize the propensity data,
\begin{equation}
Y_i = \frac{p_i - \mean{p}}{\sqrt{\mathrm{Var}[p]}},
\end{equation}
where $\mean{\cdot}$ and $\mathrm{Var}[\cdot]$ denote the mean value and the variance, respectively.
This normalization ensures that $Y_i$ has zero mean and a standard deviation of one. 

To incorporate the dynamic data for $N_\mathcal{S}$ particles, possibly taken from several different configurations, we will use a vector notation, 
\begin{equation}
    {\bf Y} = \left[ Y_1, Y_2, \dots, Y_{N_\mathcal{S}} \right]^T ,    \label{eq:Y_vector}
\end{equation}
where the superscript $T$ is the transpose operation.

Each structural feature of a given descriptor is also normalized to have zero mean and unit variance.
After normalization, the features form a vector ${\bf X}_i$ for particle $i$:
\begin{equation}
{\bf X}_i = \left[ X_i^{(1)}, \ X_i^{(2)}, \ \cdots, \ X_i^{(M)} \right]^T.
\label{eq:feature_X_original}    
\end{equation}
This vector serves as the structural input for the regression models.

To manage the whole dataset, it is convenient to introduce the following ($N_\mathcal{S} \times M$) matrix
\begin{equation}
\mathrm{X} =
\begin{bmatrix}
X_1^{(1)} & X_1^{(2)} & \cdots & X_1^{(M)} \\
X_2^{(1)} & X_2^{(2)} & \cdots & X_2^{(M)} \\
\vdots  & \vdots  & \ddots & \vdots  \\
X_{N_\mathcal{S}}^{(1)} & X_{N_\mathcal{S}}^{(2)} & \cdots & X_{N_\mathcal{S}}^{(M)} 
\end{bmatrix} ,
\label{eq:design_matrix}
\end{equation}
which is sometimes called the design matrix.

Using Eq.~(\ref{eq:feature_X_original}), $\mathrm{X}$ can be written as 
\begin{equation} \mathrm{X} = \left[ {\bf X}_1, {\bf X}_2,...,{\bf X}_{N_\mathcal{S}} \right]^T . 
\end{equation} 
Thus, ${\bf X}_i$ ($i=1,2,...,N_\mathcal{S}$) are the row vectors of $\mathrm{X}$. In this paper, we also use the column vectors ${\bf X}^{(f)}$ ($f=1,2,...,M$) of $\mathrm{X}$: \begin{equation} 
\mathrm{X}=[{\bf X}^{(1)}, {\bf X}^{(2)}, ..., {\bf X}^{(M)}] . 
\end{equation} 
The row and column vectors can be distinguished by the subscript and superscript.

In supervised learning studies, it is customary to train the model on one portion of the dataset and then test its performance on a different portion by computing some measure of correlation or statistical error.
While this aspect is less crucial for linear models than for complex deep neural networks, we follow the standard procedure of splitting the full dataset into training and test sets.
This is done by selecting a random subset of particles, $\mathcal{S}_\textrm{train}$, as training set, from the full dataset.
The fraction of selected particles in the training set is $x_\textrm{train}$.
The test set is defined by selecting a random subset of particles, $\mathcal{S}_\textrm{test}$, from the particles not included in $\mathcal{S}_\textrm{train}$.
The fraction of selected particles in the test set is, in general, $x_\textrm{test} \leq x_\textrm{train}$.

For the supervised learning methods studied in Sec.~\ref{sec:linear_regression} we found that the goodness-of-fit metrics converge when the number of datapoints per feature is above $\approx 30$, which for our dataset corresponds to about $x_\mathrm{train} = x_\mathrm{test} =0.2$.
In the following, we consider $x_\mathrm{train} = x_\mathrm{test} = 0.5$.
We use from 10 to 100 independent realizations of these sets to perform averages and assess the statistical accuracy of our results.
As an exception, the analysis of the principal component regression in Sec.~\ref{sec:pcr} involves the whole dataset.

\subsection{Pearson correlation coefficients}
\label{sec:pearson}

To illustrate some of key features of our datasets, we start with a simple analysis of correlations.
To quantify the linear dependence between two variables, $A$ and $B$, computed for a subset $\mathcal{S}$ of particles, we use the Pearson coefficient,
\begin{equation}
    \R{A}{B} = \frac{1}{N_\mathcal{S}} \sum_{i\in \mathcal{S}} \frac{(A_i - \mean{A})(B_i - \mean{B})}{\sqrt{\mathrm{Var}[A]\ \mathrm{Var}[B]}} ,
    \label{eq:pearson}
\end{equation}
computed for a set $\mathcal{S}$ comprising $N_\mathcal{S}$ particles.
By construction, $\R{A}{B}$ takes values in the range $-1 \leq \R{A}{B} \leq 1$.  

We first compute the Pearson coefficients $\R{X^{(f)}}{Y}$ between the dynamic propensity $\textbf{Y}$ and each structural feature $\textbf{X}^{(f)}$ of the BP descriptor.
Because of the weak dependence of the dynamic propensity on the species of the particles, see the Appendix~\ref{sec:appendix_species_dependence}, we include all the particles, irrespective of their species, in the analysis of correlations.
The Pearson coefficients are conveniently assembled in vector form
\begin{equation}
    \Rb{X}{Y} = \left[ \R{X^{(1)}}{Y}, \R{X^{(2)}}{Y}, \  \dots, \ \R{X^{(M)}}{Y} \right]^T .
    \label{eq:Pearson_Y_X_vector}
\end{equation}
Note that, because of the finite number of isoconfigurational trajectories, there is an upper bound on the Pearson coefficient between the propensity $\textbf{Y}$ and any structural feature~\cite{bapst2020unveiling}.
We used the method described in Ref.~\onlinecite{bapst2020unveiling} to compute the maximum Pearson coefficient $R_\textrm{max}$ that can be achieved with the current statistics.
We found $R_\textrm{max} \approx 0.98$, which is fully satisfactory for our purposes.

In Fig.~\ref{fig:pearson}, we show results for $\Rb{X}{Y}$ obtained using the BP descriptor.
The ordering of the features follows the logic described in Sec.~\ref{sec:behler_parrinello}.
We first separate the features into a radial and an angular sector.
Then, within each of these two sectors, the features are grouped into blocks according to the species or species pair for which the feature is computed.
We see that the correlations are generally modest ($|R|<0.5$) and spread over the whole set of structural features.
In particular, both radial and angular features can have similar correlations or anticorrelations with the dynamic propensity.
The presence of minima and maxima in each block of the radial features $G^\alpha$ can be connected to the peaks of the partial radial distribution functions, although we could not identify a straightforward interpretation.
The angular sector displays systematic effects because of the chemical composition of the particle environment.
Namely, features involving SS and LL pairs of neighbors are more strongly correlated or anti-correlated to the propensity.
By contrast, $R$ shows no clear trend within each of the blocks of the angular sector.

The results for the two physically motivated descriptors introduced in Sec.~\ref{sec:physical_descriptors} display qualitatively similar trends, see the Appendix~\ref{sec:appendix_datasets}.
Among all the individual structural features entering our datasets, the steric order parameter $\overline{\Theta}_i(\ell)$, coarse-grained at $\ell = 2.0$, exhibits the strongest correlation ($R \approx 0.65$) with the dynamic propensity, making it the best-performing single feature within this simple correlation analysis.
While the local volume fraction $\overline{\varphi}_i(\ell)$ and the coordination number $\overline{z}_i(\ell)$ are negatively correlated with the dynamic propensity, as expected, the local number density $\overline{\rho}_i(\ell)$ is positively correlated to it.
This counter-intuitive result is likely a nontrivial effect of composition fluctuations in our ternary glass model.

\begin{figure}[!htp]
\includegraphics[width=\linewidth]{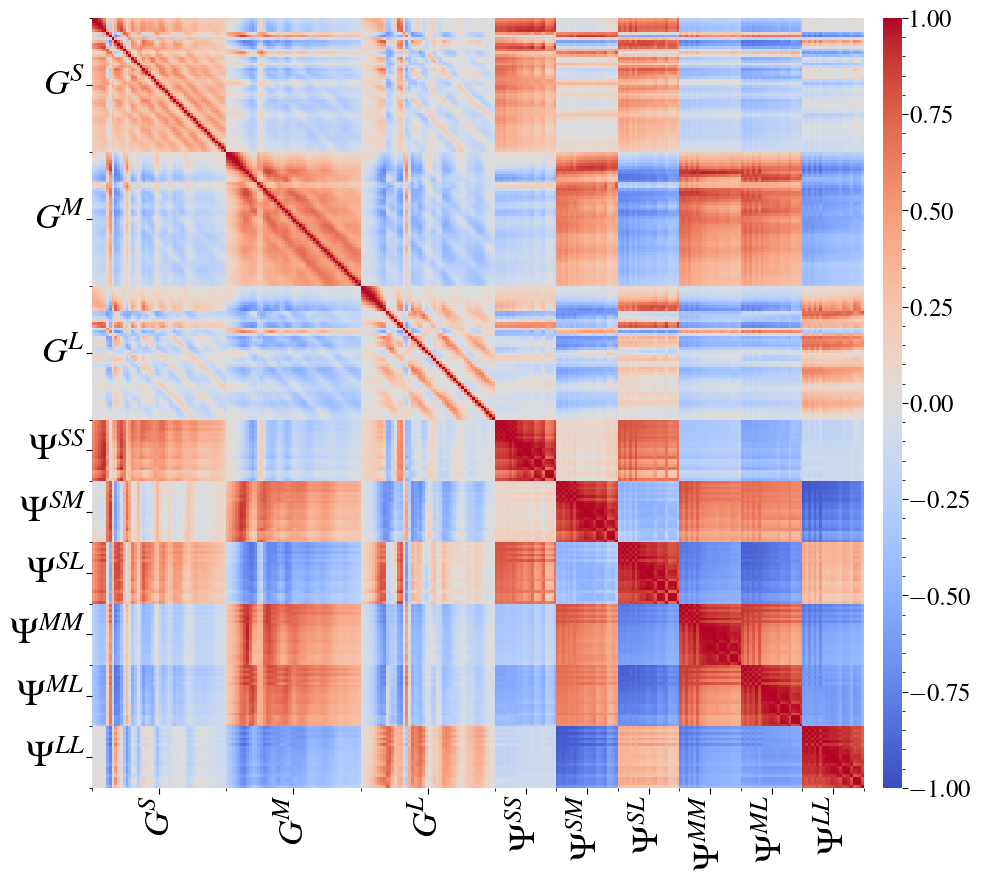}  
\caption{ 
Correlation matrix $\mathrm{C}$ for the BP descriptor. The matrix elements are given by the Pearson correlation coefficient $\R{X^{(f)}}{X^{(f')}}$.
}
\label{fig:correlation_matrix}
\end{figure}

\begin{figure*}[!ht]
\includegraphics[width=\linewidth]{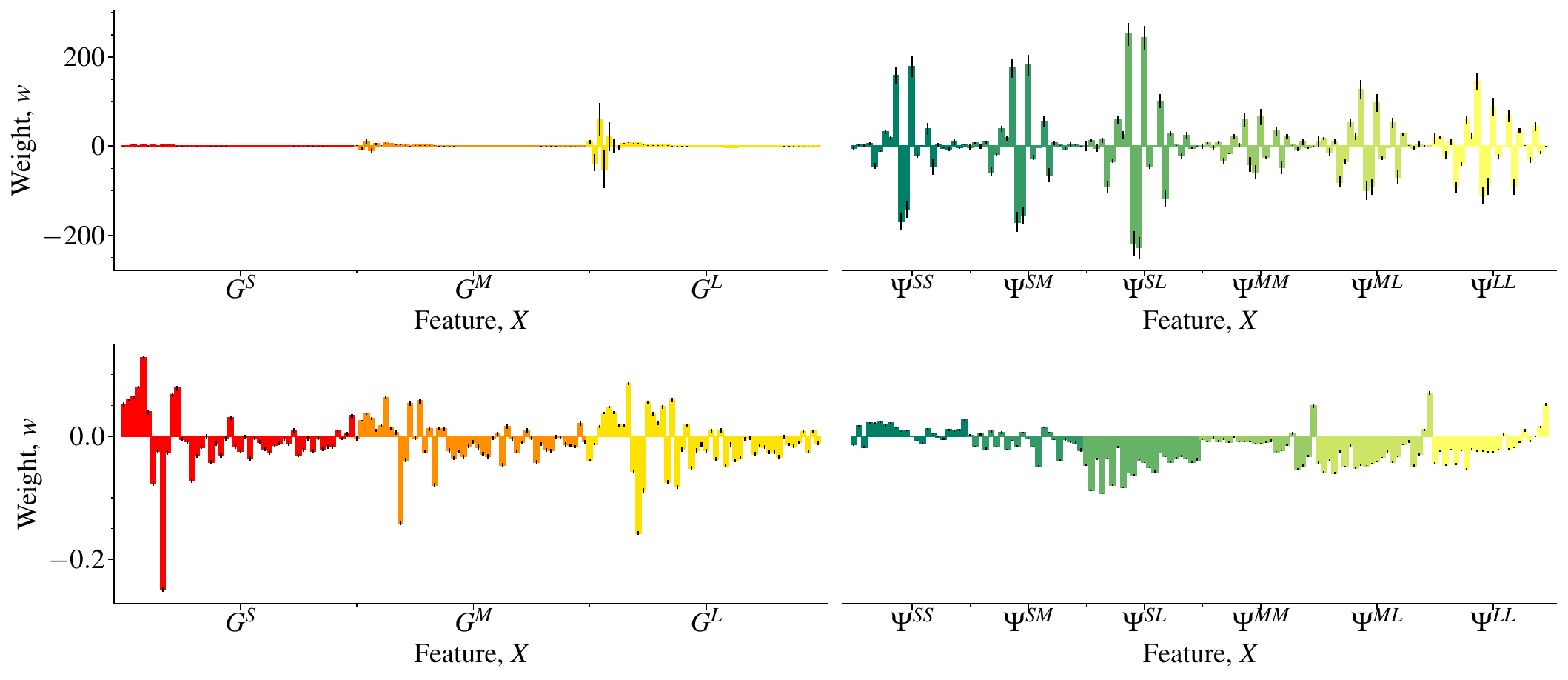}
\caption{Weights obtained from OLS and Ridge regression of the dynamic propensity using the BP descriptor:
(a) $\hat{\bf w}_\textrm{OLS}$ and (b) $\hat{\bf w}_\textrm{ridge}$ for $\alpha=10^{-1}$. The error bar corresponds to the standard deviation estimated over independent random training sets.
}
\label{fig:ridge_weights}
\end{figure*}

A common feature of all the above descriptors is the presence of significant cross-correlations between groups of structural features.
To quantify this effect, we introduce a correlation matrix, $\mathrm{C}$, whose elements are given by
\begin{equation}
    \mathrm{C}_{f, f'} = \R{X^{(f)}}{X^{(f')}} .
    \label{eq:correlation_matrix}
\end{equation}
In Fig.~\ref{fig:correlation_matrix}, we show the correlation matrix for the BP descriptor.
It has a distinct block structure, displaying strong positive or negative correlations within subsets of features.
Correlations within species-wise blocks are particularly pronounced for the angular features, see the bottom right region of Fig.~\ref{fig:correlation_matrix}.
Moreover, there are non-trivial cross-correlations also between radial and angular features.
A similar block structure is also apparent in the physically motivated descriptors, see the Appendix~\ref{sec:appendix_datasets}.
In those cases, however, block correlations are due to coarse-graining a given structural feature over a range of similar distances $\ell$.
As we will demonstrate in Sec.~\ref{sec:linear_regression}, the redundancy of these structural descriptors has a significant negative impact 
on the linear modeling of the dynamic propensity.

\section{Least square regression}
\label{sec:linear_regression}

In this section, we introduce the simplest linear regression model to describe the dynamic propensity ${\bf Y}$ using a structural descriptor $\mathrm{X}$.
The model yields a prediction of the dynamic propensity of the $i$-th particle based on the feature vector ${\bf X}_i$,
\begin{equation}
    \hat Y_i = \hat {\bf w}^T {\bf X}_i = \sum_{f=1}^M \hat w^{(f)} X_i^{(f)} ,
    \label{eq:linear_model_original}
\end{equation}
where $\hat {\bf w} = \left[ \hat w^{(1)}, \hat w^{(2)}, ...,  \hat w^{(M)} \right]^T$ are the weights.

With the vector and matrix notations in Eqs.~\eqref{eq:Y_vector} and \eqref{eq:design_matrix}, Eq.~(\ref{eq:linear_model_original}) is rewritten as
\begin{equation}
    \hat {\bf Y} = \mathrm{X}\hat {\bf w} .
    \label{eq:prediction}
\end{equation}

We consider linear models obtained by minimizing a loss function of the form
\begin{equation}
  \mathcal{L}(\hat{\bf w}) = \mathcal{L}^{\rm MSE}(\hat{\bf w}) + \mathcal{L}^{\rm reg}(\hat {\bf w}) ,
  \label{eq:loss}
\end{equation}
where
\begin{equation}
  \mathcal{L}^{\rm MSE}(\hat {\bf w}) = \frac{1}{2N_\mathcal{S}} \sum_{i \in \mathcal{S}} (\hat{Y}_i - Y_i)^2 = \frac{1}{2 N_\mathcal{S}} ||\hat{\bf Y}-{\bf Y}||^2
  \label{eq:MSE}
\end{equation}
is the mean square error (MSE) and $\mathcal{L}^{\rm reg}$ is a regularization term. In this section, we will show results for two standard linear models~\cite{james2013introduction}, namely ordinary least squares regression (Sec.~\ref{sec:ols}) and Ridge regression (Sec.~\ref{sec:ridge}), and discuss their limitations.

\subsection{Ordinary least squares regression}
\label{sec:ols}

\subsubsection{Definition}

In ordinary least squares (OLS) regression, the weight vector $\hat {\bf w}$ is determined by minimizing the MSE, \textit{i.e.}, $\mathcal{L}=\mathcal{L}^{\rm MSE}$.
Setting $\nabla_{\hat{\bf w}} \mathcal{L}^{\rm MSE}(\hat{\bf w}) = {\bf 0}$ and remembering that all the features are normalized, the solution for the OLS regression (see the Appendix~\ref{sec:appendix_regression} for the derivation) is
\begin{equation}
    \hat{\bf w}_{\rm OLS} = \mathrm{C}^{-1} \Rb{X}{Y} ,
    \label{eq:w_OLS}
\end{equation}
where $\mathrm{C}$ is the correlation matrix, 
\begin{equation}
    \mathrm{C} = \frac{1}{N_\mathcal{S}} \mathrm{X}^T \mathrm{X} 
    \label{eq:correlation_matrix_XX}
\end{equation}
and
\begin{equation}
    \Rb{X}{Y} = \frac{1}{N_\mathcal{S}} {\mathrm{X}^T {\bf Y}}   
    \label{eq:Pearson_XY}
\end{equation}
is the vector of the Pearson coefficients.

Clearly, if all features were orthogonal to each other, $\mathrm{C} = \mathrm{I}$ ($\mathrm{I}$ is the identity matrix) and $\hat{w}_{\rm OLS}^{(f)}$ would equal the Pearson coefficient $\R{X^{(f)}}{Y}$ between the dynamic propensity and feature $f$.
Then, the importance of a feature in linear regression (namely, the weight $\hat{w}_{\rm OLS}^{(f)}$) would directly match its correlation with the propensity, as in the simple analysis of Sec.~\ref{sec:pearson}.
Because of the presence of correlations between features, however, $\mathrm{C}^{-1}$ possesses non-zero off-diagonal elements, and hence $\hat{w}_{\rm OLS}^{(f)}$ is given by a linear combination of $\R{X^{(1)}}{Y}, \ldots, \ \R{X^{(M)}}{Y}$.

\subsubsection{Oscillatory behavior of the weights}

Figure~\ref{fig:ridge_weights}(a) shows the weights $\hat{\bf w}_{\rm OLS}$ obtained for the BP descriptor.
By comparing these results with Fig.~\ref{fig:pearson}, we immediately notice that the OLS solution gives a strong weight to the angular features.
Moreover, $\hat{w}_{\rm OLS}^{(f)}$ oscillates significantly, especially in the angular sector of the descriptor: the sign often changes considerably between successive features within a block.
This consistent oscillatory behavior\footnote{The oscillations are not due to limited statistics or dataset-to-dataset fluctuations, since the error bars are relatively small compared to the typical absolute value of the largest weights.} hampers the extraction of any meaningful physical insight into the relationship between the dynamical output $Y$ and the static features.
In fact, a positive (negative) weight indicates that an increase in this feature enhances (reduces) the dynamic propensity.
It would be hard to accept that very similar features, aligned along the
x-axis in Fig.~\ref{fig:ridge_weights}, can have large opposite effects on the dynamical variable $Y$.

We anticipate from Eq.~\eqref{eq:w_OLS} that the oscillations of $\hat{w}_{\rm OLS}^{(f)}$ are related to the singularity of the matrix $\mathrm{C}$.
Indeed, the invertibility of the correlation matrix $\mathrm{C}$ and the linear dependence of features are tightly connected.  
As an extreme case, one can show 
that the columns of the matrix $\mathrm{X} = [{\bf X}^{(1)}, \ldots, {\bf X}^{(M)}]$ are linearly independent if and only if the correlation matrix $\mathrm{C}$ is invertible.  
Conversely, when some features are linearly dependent, $\mathrm{C}$ is not invertible, and hence $\hat{w}_{\rm OLS}^{(f)}$ is not uniquely determined.
The results shown in Fig.~\ref{fig:ridge_weights}(a) are representative of the instability due to strong correlations among features, a phenomenon known as multicollinearity in statistical analysis~\cite{montgomery2021introduction}. We will quantify it in detail in Sec.~\ref{sec:multicollinearity}.

Interestingly, however, the Pearson coefficient between the ground truth $Y$ and the prediction $\hat Y$ of OLS regression is very good, $\R{Y}{\hat Y}\approx 0.87$, on either the train or the test datasets.
These results represent a paradigmatic machine learning case in which the prediction accuracy is very good, but the interpretability is poor.
We confirmed that the other descriptors, namely the SLO and JBB descriptors, also demonstrate this pathological behavior.

\subsection{Ridge regression}
\label{sec:ridge}

\subsubsection{Definition}

We now turn our attention to the so-called Ridge regression method~\cite{james2013introduction}, a simple variant of least squares regression commonly used to alleviate the shortcomings of OLS.
The loss function for Ridge regression, $\mathcal{L}^{\rm Ridge}(\hat {\bf w})$, 
\begin{equation}
  \mathcal{L}^{\rm Ridge}(\hat{\bf w}) = \mathcal{L}^{\rm MSE}(\hat{\bf w}) + \frac{\alpha}{2} \sum_{f=1}^M \left(\hat{w}^{(f)}\right)^2 ,
\label{eq:loss_ridge}
\end{equation}
includes a regularization term that penalizes solutions with large weights.
The parameter $\alpha$ controls the magnitude of the regularization.
Setting $\nabla_{\hat{\bf w}} \mathcal{L}^{\rm Ridge}(\hat{\bf w}) = {\bf 0}$ yields the estimated weights (see the Appendix~\ref{sec:appendix_regression} for the derivation)
\begin{equation}
    \hat{\bf w}_{\rm Ridge} = (\mathrm{C} + \alpha \mathrm{I})^{-1} \Rb{X}{Y} .
    \label{eq:w_Ridge}
\end{equation}
Of course, when $\alpha \to 0$, $\hat{\bf w}_{\rm Ridge}$ reduces to $\hat{\bf w}_{\rm OLS}$.
On the other hand, when $\alpha \agt \lambda_{\rm max}$, where $\lambda_{\rm max}$ is the largest eigenvalue of $\mathrm{C}$, we have $\hat{\bf w}_{\rm Ridge} \approx \alpha^{-1} \Rb{X}{Y}$, 
because the $\alpha \mathrm{I}$ term dominates.
In this regime, $\hat{\bf w}_{\rm Ridge}$ cannot account for non-trivial correlations in $\mathrm{C}$ and vanishes when $\alpha \to \infty$.
Therefore, $\alpha$ should be chosen smaller than an order of $\lambda_{\rm max}$.

\begin{figure}
\includegraphics[width=\linewidth]{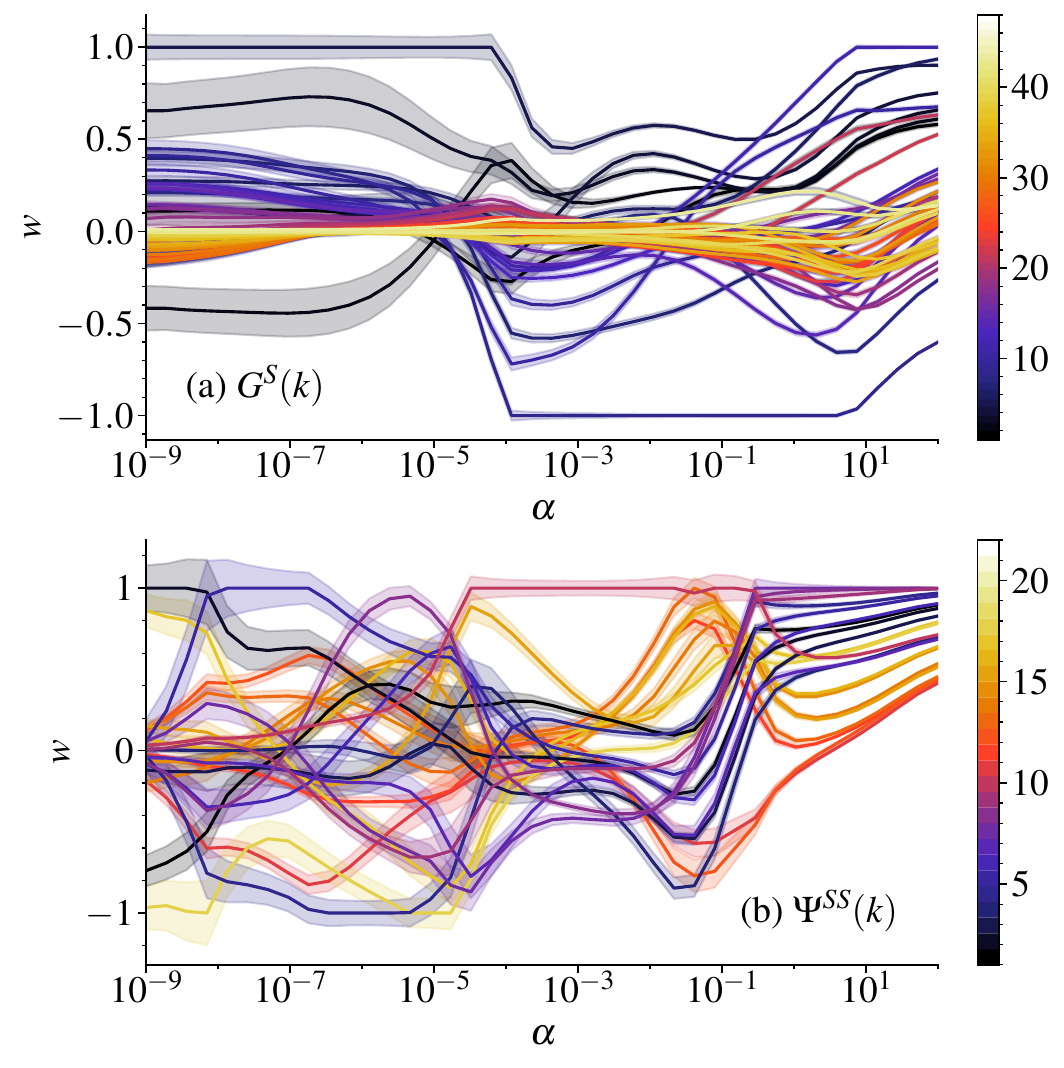}
\caption{Normalized weights as a function of the Ridge regularization parameter $\alpha$ corresponding to (a) the radial features $G^{S}(k)$ for $0\le k \le 47$ and (b) the angular features $\Psi^{SS}(k)$ for $0\le k \le 21$. The width of the shaded areas corresponds to the standard deviation estimated over independent random training sets. The color code indicates the feature index $k$.}
\label{fig:chaos}
\end{figure}

\subsubsection{Sensitivity to the regularization parameter}

To show the impact of the regularization, we show in Fig.~\ref{fig:ridge_weights}(b) $\hat{\bf w}_{\rm Ridge}$ obtained for $\alpha=0.1$.
Two main effects are observed: the amplitude of the weights is more balanced between the radial and angular sectors compared to the OLS case, and the large oscillations between successive features are suppressed.
While the main shortcoming of OLS regression seems to be solved, we notice that several features contribute with a finite weight, \textit{i.e.}, the solution is not sparse enough to be clearly interpretable.
Note that by increasing $\alpha$ further toward $\lambda_\textrm{max}\approx 100$, we find that $\hat{\bf w}_{\rm Ridge} \approx \alpha^{-1} \Rb{X}{Y}$, as anticipated above.

How sensitive are the results to the regularization parameter?
Analysis of the traces of the weights as a function of $\alpha$ provides a simple and intuitive way to assess the stability of the solutions in Ridge regression~\cite{Marquardt_Snee_1975}.
The idea is that as $\alpha$ increases from zero to small but finite values the weights will first change substantially, but there may be a range of $\alpha$ where $\hat{\bf w}_{\rm Ridge}$ they do not depend on $\alpha$ anymore.
This, of course, should occur before reaching the trivial regime where $\hat{\bf w}_{\rm Ridge} \approx \alpha^{-1} \Rb{X}{Y}$.

To illustrate the results of this analysis, we show in Fig.~\ref{fig:chaos} the Ridge traces corresponding to the features $G^{S}(k)$ and $\Psi^{SS}(k)$ in the top and bottom panels, respectively.
In each panel and for each value of $\alpha$, the weights are scaled by the largest absolute value among the weights of the corresponding subset of features.
This normalization adsorbs the huge change of scale evident from Fig.~\ref{fig:ridge_weights} and provides a vivid image of the sensitivity of the Ridge solutions.
We find that the weights are extremely sensitive to $\alpha$, especially in the angular sector of the descriptor: the traces change chaotically over a broad range of $\alpha$ spanning several orders of magnitude.
Only for $\alpha \agt 0.1$ the results seem to stabilize, in the sense that the order in the amplitude of the weights change less dramatically.
Note that the chaotic behavior of the traces is not due to limited statistics, since the estimated error bars visible in the figure are relatively small.

\subsubsection{Prediction accuracy}

Crucially, very different solutions of the regression problem, corresponding to different $\alpha$, yield predictions for the dynamic propensity with nearly identical accuracy.
To evaluate the prediction accuracy of the Ridge regression models, we use two standard metrics of performance as a function of $\alpha$. 
In addition to the Pearson coefficient, we also compute the coefficient of determination $R_2$, which focuses on the normalized total squared deviations.
If $A$ is the variable to predict, $R_2[A, \hat A]$ is given by
\begin{equation}
    R_2[A, \hat A] = 1 - \frac{\sum_{i \in \mathcal{S}} \left(A_i - \hat{A}_i\right)^2}{\sum_{i \in \mathcal{S}} \left(A_i - \mean{A} \right)^2} .
    \label{eq:coefficient_determination}
\end{equation}
When the prediction is perfect, $R_2[A, \hat A] = 1$.
Conversely, when the prediction always outputs the mean value, $R_2[A, \hat A] = 0$, which serves as a baseline.

\begin{figure}
\includegraphics[width=\linewidth]{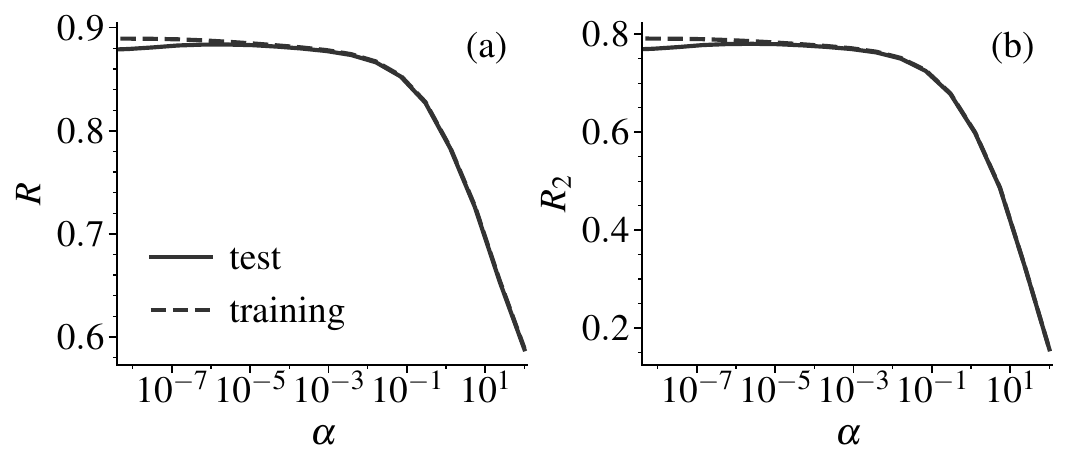}
\caption{Prediction performance metrics for Ridge regression of the dynamic propensity using the BP descriptor: (a) Pearson coefficient $R[Y, \hat{Y}]$ as a function of $\alpha$ and (b) coefficient of determination $R_2[Y, \hat{Y}]$ as a function of $\alpha$. Full and dashed lines correspond to results obtained using the test set and the train set, respectively. 
}
\label{fig:ridge_scores}
\end{figure}

Figure~\ref{fig:ridge_scores} displays the performance metrics for the BP descriptor.
We show results obtained on both the test set or the train set, to verify the lack of overfitting.
Both metrics indicate very good performance accuracy and display a flat maximum stretching over several orders of magnitudes in $\alpha$.
The performance starts to degrade appreciably only when $\alpha>0.1$, as can be seen from the similar drops of $R$ and $R_2$.
We found quantitatively similar results for the physically motivated descriptors introduced in Sec.~\ref{sec:physical_descriptors} (not shown).
The comparison between the chaotic traces of Fig.~\ref{fig:chaos} and the features performance metrics of Fig.~\ref{fig:ridge_scores} is striking.
Since very different solutions (see Fig.~\ref{fig:chaos}) yield nearly identical prediction accuracies (see Fig.~\ref{fig:ridge_scores}), it is clear that prediction alone cannot be taken as a criterion to choose $\alpha$.
This delicate aspect was overlooked in previous studies~\cite{boattini2021averaging, alkemade2022comparing, rocks2021learning}, where the regularization parameter $\alpha$ was chosen to maximize the correlation.
These issues are a manifestation of multicollinearity in the dataset and leave us with the question how to define an optimal model.

\section{Multicollinearity and its resolution}
\label{sec:multicollinearity}

In this section, we introduce a simple metric that quantifies multicollinearity in the context of linear regression, namely the condition number.
We first illustrate its qualitative behavior with a schematic two-features model (Sec.~\ref{sec:condition_number}) and then use it to revisit the linear regression models for the dynamic propensity, including Ridge regression (Sec.~\ref{sec:regression_revisited}).
The output of this analysis provides us with a range of optimal weights for Ridge regression, solving the issue of multicollinearity in a statistical sense~\cite{montgomery2021introduction}.

\subsection{Condition number}
\label{sec:condition_number}

\subsubsection{Definition}

We start by quantifying the degree of multicollinearity using the condition number $\kappa(\mathrm{C})$ of the correlation matrix $\mathrm{C}$.  
A brief introduction to the condition number and its meaning is provided in the Appendix~\ref{sec:appendix_condition_number}. 
In a nutshell, $\kappa(\mathrm{C})$ provides an upper bound for the relative error in the solution of a linear problem when small statistical perturbations are present in the data.
When $\kappa(\mathrm{C})$ is small, $\mathrm{C}$ is well-conditioned and the solution is stable. 
When $\kappa(\mathrm{C})$ is large, the matrix $\mathrm{C}$ is ill-conditioned and the error in the solution may become significant.

Since $\mathrm{C}$ is a positive semi-definite matrix, $\kappa(\mathrm{C})$ can be defined as  
\begin{equation}
    \kappa(\mathrm{C}) = \frac{\lambda_{\rm max}}{\lambda_{\rm min}} ,
\end{equation}  
where $\lambda_{\rm max}$ and $\lambda_{\rm min}$ are the largest and smallest eigenvalues of $\mathrm{C}$, respectively.  
The instability of the matrix $\mathrm{C}$ is then identified by the divergence of its condition number.
The largest eigenvalue is bounded as $\lambda_{\rm max} \leq M$, because ${\rm Tr}(\mathrm{C})=\sum_{f=1}^M \lambda^{(f)}=M$, where $\lambda^{(f)}$ are the eigenvalues.
Therefore, the divergent behavior of $\kappa(\mathrm{C})$ arises from the vanishing of the smallest eigenvalue, $\lambda_{\rm min} \to 0$.

\subsubsection{Two-features model}
\label{eq:two_features_model_condition_number}

To understand the qualitative behavior of the condition number, we consider a simple toy model composed of only two features, $\mathrm{X}=[{\bf X}^{(1)}, {\bf X}^{(2)}]$.  
The correlation matrix $\mathrm{C}$ is given by  
\begin{equation}
    \mathrm{C} = 
    \begin{bmatrix}
1 & r \\
r & 1 
\end{bmatrix},
\label{eq:C_two_features}
\end{equation}  
where $r=\R{X^{(1)}}{X^{(2)}}$ is the Pearson coefficient between the two features. 
When $r=0$, the two features are orthogonal, and $\mathrm{C}$ reduces to the identity matrix.  
When $r \to 1$ ($r \to -1$), the two features are perfectly correlated (anticorrelated), and hence ${\bf X}^{(1)}$ and ${\bf X}^{(2)}$ are linearly dependent.

Since we are interested in situations near the instability, we restrict ourselves to $0 < r \leq 1$.  
In this setting, the eigenvalues of $\mathrm{C}$ are given by $\lambda_{\rm max}=1+r$ and $\lambda_{\rm min}=1-r$.  
Therefore, the condition number is  
\begin{equation}
    \kappa(\mathrm{C}) = \frac{1+r}{1-r}.
\end{equation}  
When $r \to 1$, $\lambda_{\rm min} \to 0$ and hence the condition number diverges.  
Simultaneously, $\mathrm{C}$ becomes non-invertible since the determinant vanishes. 
Thus, $\kappa(\mathrm{C})$ effectively signals the presence of multicollinearity.

\subsubsection{Origin of the oscillatory behavior of weights}

We now seek to explain the oscillatory behavior of the weights observed in Fig.~\ref{fig:ridge_weights}. This behavior arises from the ill-conditioning of the correlation matrix, because of 
the vanishing of its smallest eigenvalue $\lambda_{\rm min}$.

The key observation is that two strongly correlated features tend to acquire weights of opposite signs with large magnitude, even though such features are expected to be physically similar. 
This hampers the physical interpretation of the weights, which should reflect the importance of features contributing to the dynamics.
The essence of this problem can be reduced to a two-feature model.
Suppose that $\Rb{X}{Y} = [R^{(1)}, R^{(2)}]^T$, where $R^{(1)}$ and $R^{(2)}$ are some values with $-1 \leq R^{(1)}, R^{(2)} \leq 1$.  
Using Eq.~(\ref{eq:w_OLS}), the estimated weights $\hat{\bf w}_{\rm OLS} = [\hat w_{\rm OLS}^{(1)}, \hat w_{\rm OLS}^{(2)}]^T$ are given by
\begin{eqnarray}
\hat w_{\rm OLS}^{(1)} &=& \frac{1}{\lambda_{\rm max}\lambda_{\rm min}} \left( R^{(1)} - r R^{(2)} \right), \\
\hat w_{\rm OLS}^{(2)} &=& - \frac{1}{\lambda_{\rm max}\lambda_{\rm min}} \left( rR^{(1)} - R^{(2)} \right).
\end{eqnarray}
When $r \simeq 1$, the magnitude of both weights becomes very large due to $\lambda_{\rm min} \to 0$, and the signs become opposite, $\hat w_{\rm OLS}^{(1)} \simeq -\hat w_{\rm OLS}^{(2)}$. 

\subsection{Least squares regression revisited}
\label{sec:regression_revisited}

We now turn our attention to the structural dataset given by the BP descriptor and quantify its degree of multicollinearity.
A direct calculation of the condition number yields $\kappa(\mathrm{C}) \approx 1.4 \times 10^{18}$.
As a rule of thumb~\cite{montgomery2021introduction}, condition numbers smaller than 100 are not problematic, while values greater than 1000 indicate problems with multicollinearity.
We found similar orders of magnitude for the structural datasets based on the physically motivated descriptors, \textit{i.e.}, $4.5\times 10^{17}$ and $1.2\times 10^{15}$ for the JBB and SLO descriptors, respectively.
Also the original BP descriptor used in Refs.~\cite{cubuk2015identifying, rocks2021learning}, in which features were not coarse-grained, yields a condition number of order $10^{17}$.
We conclude that several structural datasets recently used in several recent glassy materials studies~\cite{cubuk2015identifying, rocks2021learning, jung2023predicting, sharma2024selecting} are severely affected by multicollinearity, much more than typical datasets analyzed in statistics textbooks~\cite{montgomery2021introduction}.

\begin{figure}
\includegraphics[width=\linewidth]{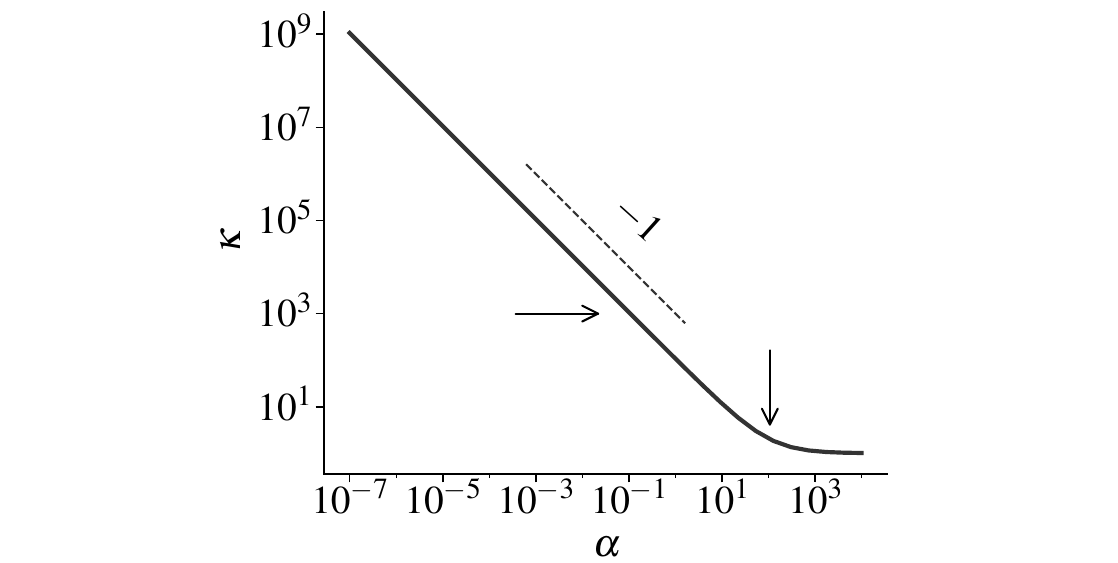}
\caption{ 
Condition number $\kappa(\mathrm{C} + \alpha \mathrm{I})$ as a function of the Ridge regularization parameter $\alpha$. The vertical arrow marks the largest eigenvalue of $\mathrm{C}$. The horizontal arrow is drawn at a condition number equal to 1000. The dashed line indicates an inverse power law.}
\label{fig:multicollinearity}
\end{figure}

We can now build on the multicollinearity analysis to identify a range of optimal solutions within Ridge regression.
To see how the regularization term in Eq.~\eqref{eq:loss_ridge} mitigates the effect of multicollinearity, we compute the condition number in the Ridge setting.
We get 
\begin{equation}
    \kappa(\mathrm{C} + \alpha \mathrm{I}) = \frac{\lambda_{\rm max} + \alpha}{\lambda_{\rm min} + \alpha},
    \label{eq:kappa_alpha}
\end{equation}
which avoids the divergence of $\kappa(\mathrm{C} + \alpha \mathrm{I})$ when multicollinearity is severe ($\lambda_{\rm min} \to 0$).
In Fig.~\ref{fig:multicollinearity}, we show $\kappa(\mathrm{C} + \alpha \mathrm{I})$ as a function of $\alpha$.
In the range $\lambda_\textrm{min} \ll \alpha \ll \lambda_\textrm{max}$, the condition numbers decreases like $\lambda_\textrm{max} \alpha^{-1}$.
To reduce the effects of multicollinearity to an acceptable degree, one should lower the condition numbers down to at least $10^3$ or less~\cite{montgomery2021introduction}.
This occurs around $\alpha = 0.1$. 
The regularization becomes meaningless for $\alpha \agt \lambda_{\rm max}$, since it entirely suppresses correlations in $\mathrm{C}$.
The crossover of $\kappa$ towards 1 occurs when $\alpha$ becomes of the order of $\lambda_\textrm{max} \approx 100$, as confirmed by Fig.~\ref{fig:multicollinearity}.
Since values of $\alpha$ in the range $0.1 \alt \alpha \alt 100$ are acceptable but the performance accuracy decreases systematically in this range, see Fig.~\ref{fig:ridge_scores}, we tentatively choose $\alpha=0.1$ as an optimal value.
Finally, one can also see how the oscillatory behavior originating from multicollinearity is suppressed in Ridge regression.
This point can already be grasped by comparing Fig.~\ref{fig:ridge_weights}(a) and Fig.~\ref{fig:ridge_weights}(b), but is best understood in the principal component basis, see the Appendix~\ref{sec:oscillation_Ridge_PCA}.

Before closing this section, we note that, in addition to the condition number, we also computed the variance inflation factor (VIF)~\cite{montgomery2021introduction}, another popular metric for evaluating the degree of multicollinearity. The VIF acts as a scaling factor for the variance of the estimated weights arising from dataset-to-dataset fluctuations, and it can diverge under severe multicollinearity even when the sample size is sufficiently large. We find that the VIF provides results consistent with the condition number, when applying the usual rules of
thumb~\cite{montgomery2021introduction}. However, this quantity does not offer additional insight for our dataset, because the sample size is quite large (and thus the estimated error bars in Fig.~\ref{fig:ridge_weights} are small). For this reason, we do not report these results in this paper and leave their investigation to future work, particularly in situations with limited sample sizes.

\section{Towards interpretable linear regression models}
\label{sec:resolutions}

The analysis presented in Sec.~\ref{sec:multicollinearity} shows that Ridge regression is an effective method to cope with the effects of multicollinearity in structure-dynamics datasets.
One could then identify the most relevant features as those with the largest absolute weights, see Sec.~\ref{sec:discussion}.
However, Ridge regression does not offer a principled way to perform feature selection: by definition, the model retains all the features.
To achieve a simple physical picture, we must substantially reduce the dimensionality of the problem at hand.
In this section, we consider a generalization of Ridge regression, called elastic net~\cite{zou2005regularization}, that provides means to select the most relevant features, while reducing multicollinearity (Sec.~\ref{sec:elastic_net}).
As an alternative approach, we use principal component regression, which builds upon a simple linear transformation of original features (Sec.~\ref{sec:pcr}).

\subsection{Elastic net and Lasso regression}
\label{sec:elastic_net}

\subsubsection{Definition}

The elastic net model is an extension of Ridge regression to perform feature selection~\cite{zou2005regularization}.
In this method, the regularization term in Eq.~\eqref{eq:loss} reads

\begin{equation}
  \mathcal{L}^{\rm reg}(\hat{\bf w}) =  a \sum_{f=1}^M \left|\hat{w}^{(f)}\right| + \frac{b}{2} \sum_{f=1}^M \left(\hat{w}^{(f)}\right)^2 ,
  \label{eq:loss_elastic_net}
\end{equation}
where $a$ and $b$ are two regularization parameters.
In addition to the L2 norm term already included in Ridge regression, Eq.~\eqref{eq:loss_elastic_net} now includes an L1 norm regularization term.
Like in Lasso regression~\cite{james2013introduction}, the L1 term shrinks some of the weights to zero within numerical accuracy to perform feature selection.
Following the elastic net implementation of \texttt{scikit-learn}~\cite{pedregosa2011scikit}, we define $\alpha=a+b$ and $\beta=a/(a+b)$, where now $0 \le \beta \le 1$.
$\alpha$ quantifies the overall strength of regularization, incorporating both L1 and L2 terms, while $\beta$ controls the relative contribution of L1 regularization.
$\beta = 0$ and $\beta = 1$ correspond to pure Ridge and Lasso regression, respectively.
For any pair $(\alpha$, $\beta)$, the method finds an optimal number of features, $P$, that minimizes the loss function.

\begin{figure}
\includegraphics[width=\linewidth]{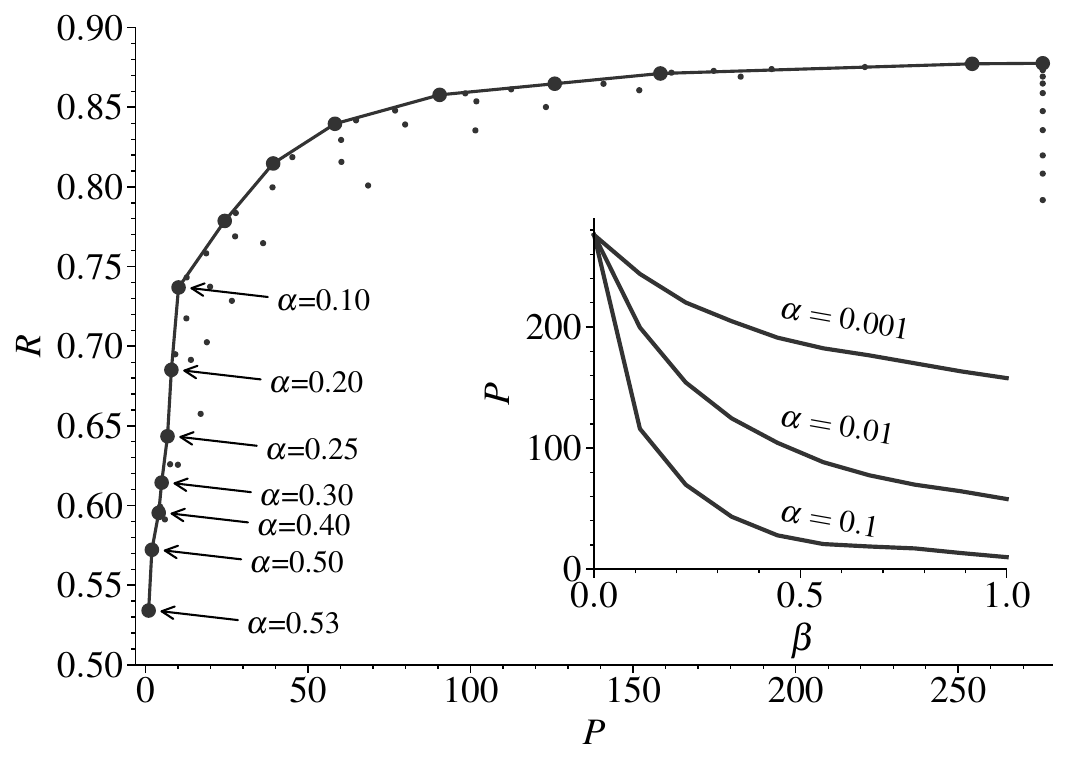}
\caption{Elastic net regression of the dynamic propensity using the BP descriptor. Main panel: Pearson coefficient $R[Y,\hat{Y}]$ for the Lasso regression models (large circles) as a function of the number of selected features $P$. The values of $\alpha$ corresponding to the models with $P\le 10$ are indicated in the figure. The small circles indicate results for values $(\alpha, \beta)$ with $\beta<1$. Inset: number of selected features $P$ as a function of $\beta$ for selected values of $\alpha$.}
\label{fig:elastic_net}
\end{figure}

\begin{table*}[!hbt]
  \centering
    {\renewcommand{\arraystretch}{1.1} \begin{tabular}{lcccccccccccccc}
\hline
\hline
&&&$\Psi^{SS}(20)$&$\Psi^{SS}(19)$&$G^S(11)$&$G^S(5)$&$G^S(4)$&$\Psi^{SM}(18)$&$G^M(11)$&$G^L(12)$&$G^M(1)$&$G^M(9)$&$G^S(8)$&$G^L(18)$\\[0pt]
\cline{4-15}
$\alpha$&$P$&$R$&\textit{0.53}&\textit{0.53}&\textit{0.53}&\textit{0.51}&\textit{0.47}&\textit{0.39}&\textit{0.35}&\textit{0.31}&\textit{0.28}&\textit{-0.20}&\textit{-0.28}&\textit{-0.42}\\[0pt]
\hline
0.1&10&$0.74$&$\bullet$&&$\bullet$&&$\bullet$&$\bullet$&$\bullet$&$\bullet$&$\bullet$&$\bullet$&$\bullet$&$\bullet$\\[0pt]
0.2&8&$0.69$&$\bullet$&&$\bullet$&$\bullet$&&$\bullet$&$\bullet$&$\bullet$&&&$\bullet$&$\bullet$\\[0pt]
0.25&7&$0.64$&$\bullet$&$\bullet$&$\bullet$&$\bullet$&&$\bullet$&&$\bullet$&&&&$\bullet$\\[0pt]
0.3&5&$0.61$&$\bullet$&$\bullet$&$\bullet$&$\bullet$&&&&&&&&$\bullet$\\[0pt]
0.4&4&$0.60$&$\bullet$&$\bullet$&$\bullet$&$\bullet$&&&&&&&&\\[0pt]
0.45&3&$0.59$&$\bullet$&$\bullet$&$\bullet$&&&&&&&&&\\[0pt]
0.5&2&$0.57$&$\bullet$&&$\bullet$&&&&&&&&&\\[0pt]
0.53&1&$0.53$&$\bullet$&&&&&&&&&&&\\[0pt]
\hline
\hline
\end{tabular}}

\caption{\label{table:elastic_net} Low-dimensional optimal models obtained using Lasso regression with the BP descriptor. The bullets indicate the most probable features selected by the models for selected values of $\alpha$. The numbers in italic below each feature are the Pearson coefficients between that feature and the dynamic propensity. The estimated statistical uncertainty on the Pearson coefficients is $\pm 0.01$.}
\end{table*}

\subsubsection{Model selection}

To get a feeling for the role of $\beta$, we show in the inset of Fig.~\ref{fig:elastic_net} the dependence of the number of selected features $P$ on $\beta$, for a few fixed values of $\alpha$.
We average the results over five independent realizations of the train and test sets.
As expected, the number of selected features decreases  as the strength of the L1 regularization increases, reaching its minimum for $\beta=1$.
A similar effect is observed with increasing $\alpha$.
Note that a too large value of $\alpha$ may lead to the trivial solution $P=0$.
For our dataset, we found that Lasso regression ($\beta=1$) requires $\alpha < 0.53$ to yield a finite $P$.

In general, the prediction performance of the model will vary with $\alpha$ and $\beta$.
The main panel of Fig.~\ref{fig:elastic_net} shows the Pearson coefficient $R[Y,\hat{Y}]$ between the solutions of several elastic net models and the dynamic propensity, as a function of $P$.
The data for $\beta=1$ are indicated as large black symbols.
We also include the results of a more extensive sampling of $(\alpha, \beta)$ pairs, shown as small dots.
The envelope of $R$ vs. $P$ defines a set of optimal models, in the sense that they provide the best performance accuracy for a given size $P$ of the feature space.
It is clear from the figure that these models coincide with those identified by Lasso regression.
The typical values of the Pearson coefficients for the low-dimensional optimal models range from 0.53 for $P=1$, to 0.61 for $P=5$, to 0.74 for $P=10$.
These values can be compared with the maximum value of 0.88 achieved when retaining the full descriptor, see also Sec.~\ref{sec:discussion}.

In Table~\ref{table:elastic_net}, we report the features selected by low-dimensional optimal models ($P\le 10$) within Lasso regression.
Note that these are the most probable selected features, estimated using 10 independent realizations of the train and test sets.
We include the corresponding average Pearson coefficients with the dynamic propensity, both for the regression models and for the individual features.
The low-dimensional models of Lasso regression successfully pinpoint the features that are most correlated to the dynamic propensity, found in either angular or radial sectors of the BP descriptor.
While these models achieve a reasonable correlation with the dynamic propensity, they are still somewhat redundant.
For instance, the features selected by the $P=2$ model, \textit{i.e.}, $\Psi^{SS}(20)$ and $G^S(11)$, are appreciably correlated with one another ($R=0.62$).
Even worse, the models with $P=3, 4, 5$ and 7 contain two nearly identical angular features, $\Psi^{SS}(19)$ and $\Psi^{SS}(20)$, having mutual correlation $R=0.99$.
This can be explained by the fact that for these models $\beta=1$, hence there is no suppression of multicollinearity.
It may be possible to address this issue by using a minimum-redundancy-maximum-relevance selection scheme~\cite{peng2005feature}, which includes iteratively features that are highly correlated with the target, while penalizing those strongly correlated with the ones already taken.
We leave an assessment of this approach to a future investigation.

\subsection{Principal component regression}
\label{sec:pcr}

\subsubsection{Definition}

In this section, we use principal component analysis (PCA) as an alternative approach to reducing the dimensionality of the problem.
This method allows one to find a new set of relevant features, defined in an orthogonal basis formed by the eigenvectors of the correlation matrix $\mathrm{C}$.
The corresponding modes are called principal components and can be used to perform linear regression.
When only a subset of relevant principal components is selected, this regression model is called principal component regression (PCR).

Let us briefly remind the key ingredients of PCA. 
Since $\mathrm{C}$ is a symmetric matrix, it can be diagonalized using the orthogonal matrix $\mathrm{U} = [{\bf u}^{(1)}, {\bf u}^{(2)}, \dots, {\bf u}^{(M)}]$, where ${\bf u}^{(f)}$ are the eigenvectors of $\mathrm{C}$, and $\lambda^{(f)}$ are the corresponding eigenvalues. This gives the equation
\begin{equation}
    \mathrm{C} = \mathrm{U} \mathrm{\Lambda} \mathrm{U}^T ,
    \label{eq:PCA_diagonal}
\end{equation}
where $\mathrm{\Lambda}$ is a diagonal matrix with eigenvalues $\lambda^{(1)}, \dots, \lambda^{(M)}$, sorted as $\lambda^{(1)} \geq \dots \geq \lambda^{(M)} > 0$.

We now obtain new features $\mathrm{X'}$ through a linear transformation, $\mathrm{X'} = \mathrm{XU}$. 
The original feature ${\bf X}_i$ in Eq.~(\ref{eq:feature_X_original}) is expanded using the orthogonal basis ${\bf u}^{(1)}, {\bf u}^{(2)}, \dots, {\bf u}^{(M)}$, and the coefficients ${X'}_i^{(f)}$ represent the new features on this basis. Specifically, we have
\begin{eqnarray}
  {\bf X}_i &=& \sum_{f=1}^M {X'}_i^{(f)}{\bf u}^{(f)}, \\
  {X'}_i^{(f)} &=& \left({\bf u}^{(f)}\right)^T {\bf X}_i .
                   \label{eq:X_prime}
\end{eqnarray}

\begin{figure}
\includegraphics[width=\linewidth]{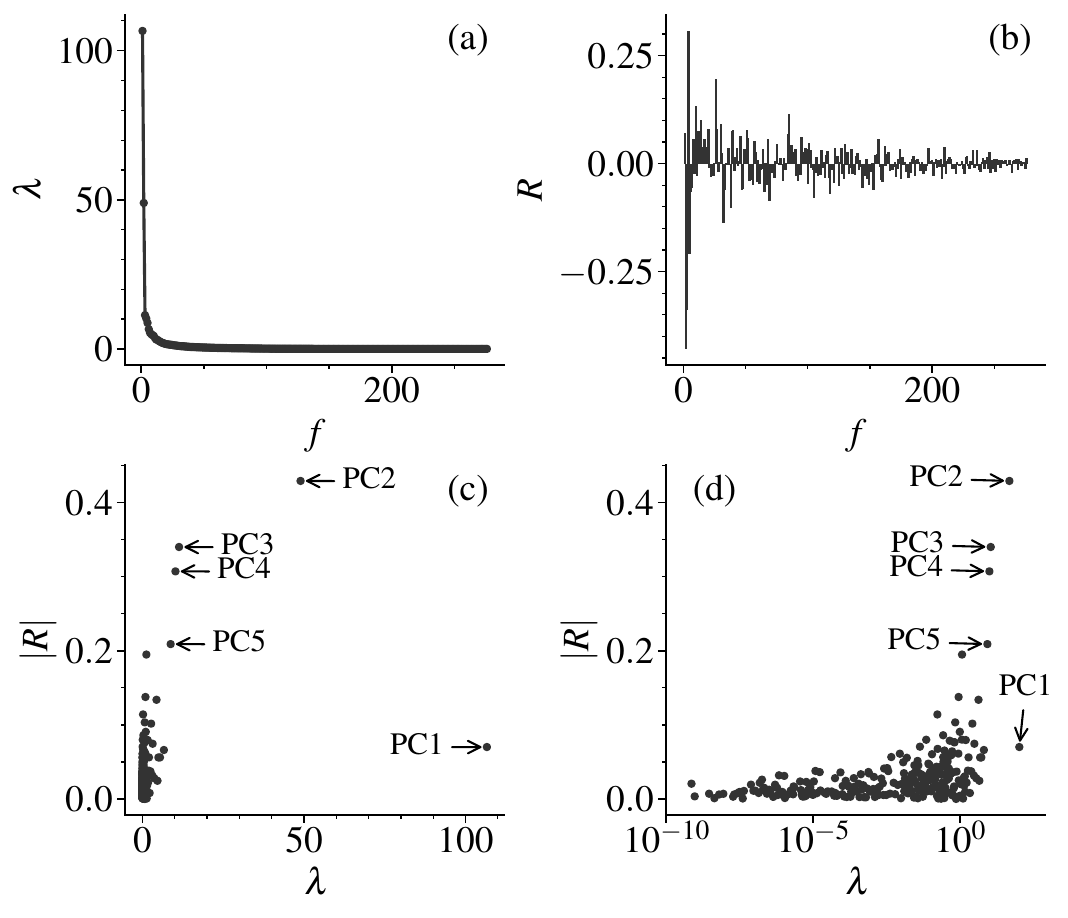}
\caption{Principal component analysis for the BP descriptor.
  (a) Eigenvalue $\lambda^{(f)}$ of each PC; (b) Pearson coefficient $R[\tilde{X}^{(f)}, Y]$ between the dynamic propensity and each PC; (c) absolute value of the Pearson coefficient $|R[\tilde{X}^{(f)}, Y]|$ between the dynamic propensity and each PC as a function of the eigenvalue $\lambda^{(f)}$ of the PC; (d) same as (c) but showing the eigenvalues in log scale. In panels (c) and (d), the top 5 PCs are marked by arrows.
}
\label{fig:pca}
\end{figure}

From Eq.~\eqref{eq:X_prime}, we see that the mean value of ${X'}_i^{(f)}$ is zero. The covariance matrix for $\mathrm{X'}$ is given by
\begin{equation}
\frac{1}{N_\mathcal{S}}{\mathrm{X}'}^T \mathrm{X}' = \mathrm{U}^T \mathrm{C} \mathrm{U} = \mathrm{\Lambda},
\end{equation}
hence the variance of ${X'}_i^{(f)}$ is $\lambda^{(f)}$.
We show $\lambda^{(f)}$ in Fig.~\ref{fig:pca}(a).
The first two PCs have very large eigenvalues, while the remaining components form a tail that is difficult to appreciate on a linear scale.

Let us now normalize the features
\begin{equation}
    \tilde X_i^{(f)} = \frac{{X'}_i^{(f)}}{\sqrt{\lambda^{(f)}}} ,
    \label{eq:def_new_features}
\end{equation}
so that the new feature $\tilde X_i^{(f)}$ has zero mean and unit variance, and determine how $\Rb{X}{Y}$ is transformed in the PC basis.
Then, by applying $\mathrm{U}^T$, we obtain
\begin{equation}
    \mathrm{U}^T \Rb{X}{Y} = \frac{1}{N_\mathcal{S}} \mathrm{X'}^T {\bf Y} = \mathrm{\Sigma} \Rb{\tilde{\textbf{X}}}{Y} ,
    \label{eq:U_rho_YX}
\end{equation}
where $\mathrm{\Sigma}$ (not to be confused with summation) is a diagonal matrix whose elements are $\sqrt{\lambda^{(1)}}, \sqrt{\lambda^{(2)}}, \dots, \sqrt{\lambda^{(M)}}$, meaning that $\mathrm{\Sigma^2=\Lambda}$. We show $\Rb{\tilde{\textbf{X}}}{Y}$ in Fig.~\ref{fig:pca}(b).
Only the first few PCs have non-negligible correlations with the dynamic propensity, say $|R|> 0.3$.
This is best appreciated from panels (c) and (d) of Fig.~\ref{fig:pca}, where we show the absolute value of $\Rb{\tilde{\textbf{X}}}{Y}$ as a function of $\lambda^{(f)}$.
Surprisingly, PC1 has a negligible correlation with the dynamic propensity, while it is PC2 that has the largest correlation.
Thus, selecting the PCs on the basis of their eigenvalue alone, in an unsupervised fashion, can lead to suboptimal performance accuracy in a regression model.
The Pearson coefficients of the remaining PCs from $f=3$ to $f=5$ decrease with increasing $f$, and the remaining PCs form a background of uncorrelated features.

\subsubsection{Interpretation of the principal components}

In principle, inspection of the eigenvector $\textbf{u}^{(f)}$ should reveal the nature of the structural fluctuations occurring along each mode: elements of $\textbf{u}^{(f)}$ that are large in absolute value indicate the features that contribute the most to the fluctuations along a given principal component.
Unfortunately, the interpretation of the eigenvectors is not always straightforward, since they are a linear combination of all the original features: indeed, for the BP descriptor, we found that no subset of features stands out in the PCs.
Thus, one must look for statistical correlations with simpler physical variables.

\begin{figure}
\includegraphics[width=\columnwidth]{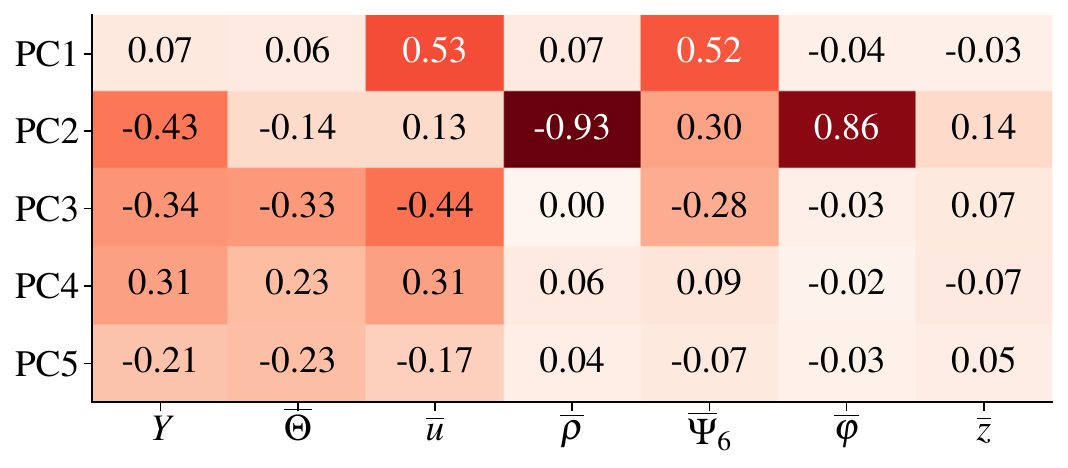}  
  \caption{Pearson coefficients $R$ between the first five PCs and the physically motivated features. The latter are coarse-grained over a length $\ell=1.5$.}
\label{fig:eigenvectors_bp}
\end{figure}

In Fig.~\ref{fig:eigenvectors_bp} we report the Pearson coefficients between the projections on the first few PCs and the physically motivated features defined in Sec.~\ref{sec:physical_descriptors}.
Fluctuations along PC1 are moderately correlated to concomitant fluctuations of the local potential energy $\overline u$ and the bond orientational order parameter $\overline \Psi_6$.
This structural mode gathers the bulk of the variance of the normalized dataset, but it is uncorrelated to the dynamic propensity.
The interpretation of the second PC is sharp: structural fluctuations along PC2 are strongly correlated to the local number density $\overline \rho$ ($|R|=0.93$).
Note that, as anticipated in Sec.~\ref{sec:pearson}, in our model the fluctuations of the local packing fraction $\overline \varphi$ are strongly anti-correlated with those of $\overline \rho$.
The remaining PCs do not seem to have clear connections with physically motivated features.
Interestingly, none of the structural modes identified by PCA captures the structural fluctuations associated to $\overline \Theta$, which is most strongly correlated with the dynamic propensity.
Note, however, a regression of $\overline \Theta$ against the full BP descriptor yields quite accurate results: the representation of this feature is simply spread over several features of the descriptor.

\subsubsection{Linear regression in the PCA basis}

We now frame the solutions of linear regression models, Eq.~\eqref{eq:prediction}, in the context of PCA.
Note that, strictly speaking, the analysis below applies only when correlations are computed on the training set, as is the case here, but it provides an informative interpretation for test data predictions as well.

Let us first focus on the OLS regression case.
As shown in the Appendix~\ref{sec:appendix_pcr}, the dynamic propensity predicted by OLS regression is
\begin{equation}
    \hat {\bf Y} = \sum_{f=1}^M \R{\tilde X^{(f)}}{Y} \tilde {\bf X}^{(f)} .
    \label{eq:Y_PCA_OLS}
\end{equation}
The coefficient in front of each feature $\tilde {\bf X}^{(f)}$, which reflects its importance in the prediction, is simply the Pearson correlation coefficient shown in Fig.~\ref{fig:pca}(b). 
This is, of course, because all the PC eigenvectors are orthogonal to each other.
The Pearson coefficient between the ground truth propensity and the predicted one is then easily found:
\begin{equation} 
  \R{Y}{\hat Y} = \sqrt{\sum_{f=1}^M (\R{\tilde X^{(f)}}{Y})^2} .
  \label{eq:pca_pearson_ols}
\end{equation} 
This expression provides a clear geometric interpretation of how each $\R{\tilde X^{(f)}}{Y}$ contributes to the performance $\R{Y}{\hat Y}$ in training set.

\subsubsection{Dimensional reduction and comparison with elastic net}

\begin{figure}
\includegraphics[width=\columnwidth]{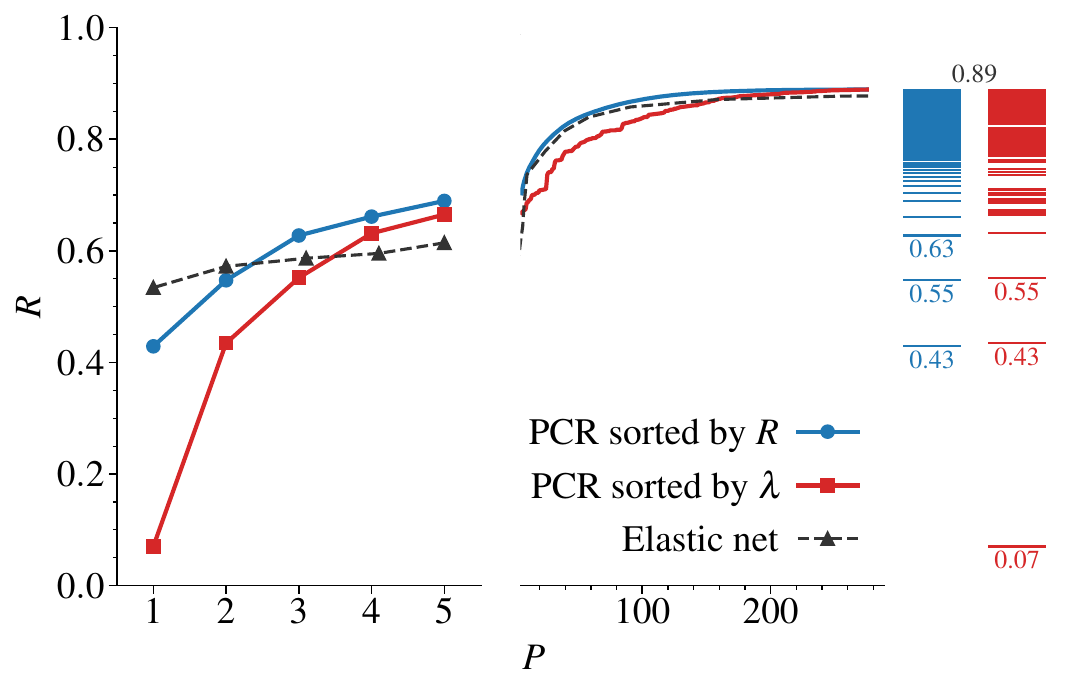}
\caption{Pearson coefficient $R$ between the ground truth dynamic propensity and the PCR  predictions using the first $P$ PCs sorted according to their eigenvalue (squares) or their Pearson coefficient with the dynamic propensity (circles). The results of the optimal elastic net regression models are also included (triangles). Note the change of scale on the $x$ axis after $P=5$ components. The tics on the right side of the figure indicate the values of the Pearson coefficient reached for a given value of number of PCs included in the regression model.
}
\label{fig:pcr}
\end{figure}

To perform dimensional reduction, one selects only $\tilde {\bf X}^{(1)}, \tilde {\bf X}^{(2)}, ..., \tilde {\bf X}^{(P)}$ with $P \ll M$, yielding
\begin{equation}
    \hat {\bf Y} = \sum_{f=1}^P \R{\tilde X^{(f)}}{Y} \tilde{\bf X}^{(f)} .
\end{equation}
This operation corresponds to neglecting features associated with smaller eigenvalues $\lambda^{(P+1)}, \dots, \lambda^{(M)}$.
In Fig.~\ref{fig:pcr}, we analyze the performance of PCR by showing $\R{Y}{\hat Y} = \sqrt{\sum_{f=1}^P (\R{\tilde X^{(f)}}{Y})^2}$ as a function of the number of components $P$.
Inspired by Eq.~\eqref{eq:pca_pearson_ols}, we also include results obtained by sorting the features according to the absolute value of $\R{\tilde X^{(f)}}{Y}$.
This corresponds to a supervised feature extraction scheme.
On the side of the figure, we include as ticks the Pearson correlation coefficients achieved with a given number of features, $P$.
The visual impression is that the first few PCs contribute to the bulk of the correlation, reaching $R=0.7$ with $P=5$.
The rest of the PCs constitute a dense, poorly interpretable background of orthogonal variables that are essentially uncorrelated with the propensity, see Fig.~\ref{fig:pca}(b).
Linear regression methods can nonetheless harvest this background to reconstruct the dynamic propensity field precisely.

It is interesting to compare these results to those obtained by the optimal elastic net models, which are included as dashed lines in Fig.~\ref{fig:pcr}.
We see that the elastic net models converge to the maximum correlation, as a function of selected features, approximately like in supervised PCR.
For small numbers of features, the two regression models provide about the same performance accuracy, although the precise details depend on the descriptor.
One advantage of PCR is that the feature extraction scheme is straightforward and efficient, thanks to the orthogonality of the PCA basis.
By contrast, although the features selected by elastic net are not orthogonal to each other, they are more easily interpretable than the PC eigenvectors.
The latter, in fact, mix all the features of the original descriptor 
and do not always lend themselves to a straightforward interpretation.
We will further discuss this issue in Sec.~\ref{sec:discussion}.

\begin{table}
  \centering
  {\renewcommand{\arraystretch}{1.1} \begin{tabular*}{\linewidth}{@{\extracolsep{\fill}} lrrrrr}
\hline
\hline
 & & \multicolumn{2}{c}{Ridge} & \multicolumn{2}{c}{PCR} \\
\cline{3-4} \cline{5-6}
&Features&$\alpha=0$&$\alpha=0.1$&$P=5$&$P=2$\\[0pt]
\hline
BP descriptor&276&0.87&0.85&0.69&0.55\\[0pt]
JBB descriptor&120&0.35&0.87&0.80&0.53\\[0pt]
SLO descriptor&60&0.86&0.85&0.84&0.81\\[0pt]
\hline
\hline
\end{tabular*}}
  
  \caption{\label{table:performance}Pearson coefficient between the dynamic propensity and selected linear regression models for all the investigated descriptors.
  The performance of the JBB descriptor without regularization ($\alpha = 0$) is very low, due to severe overfitting.}
\end{table}

\begin{figure*}[!htb]
  \includegraphics[width=0.82\linewidth]{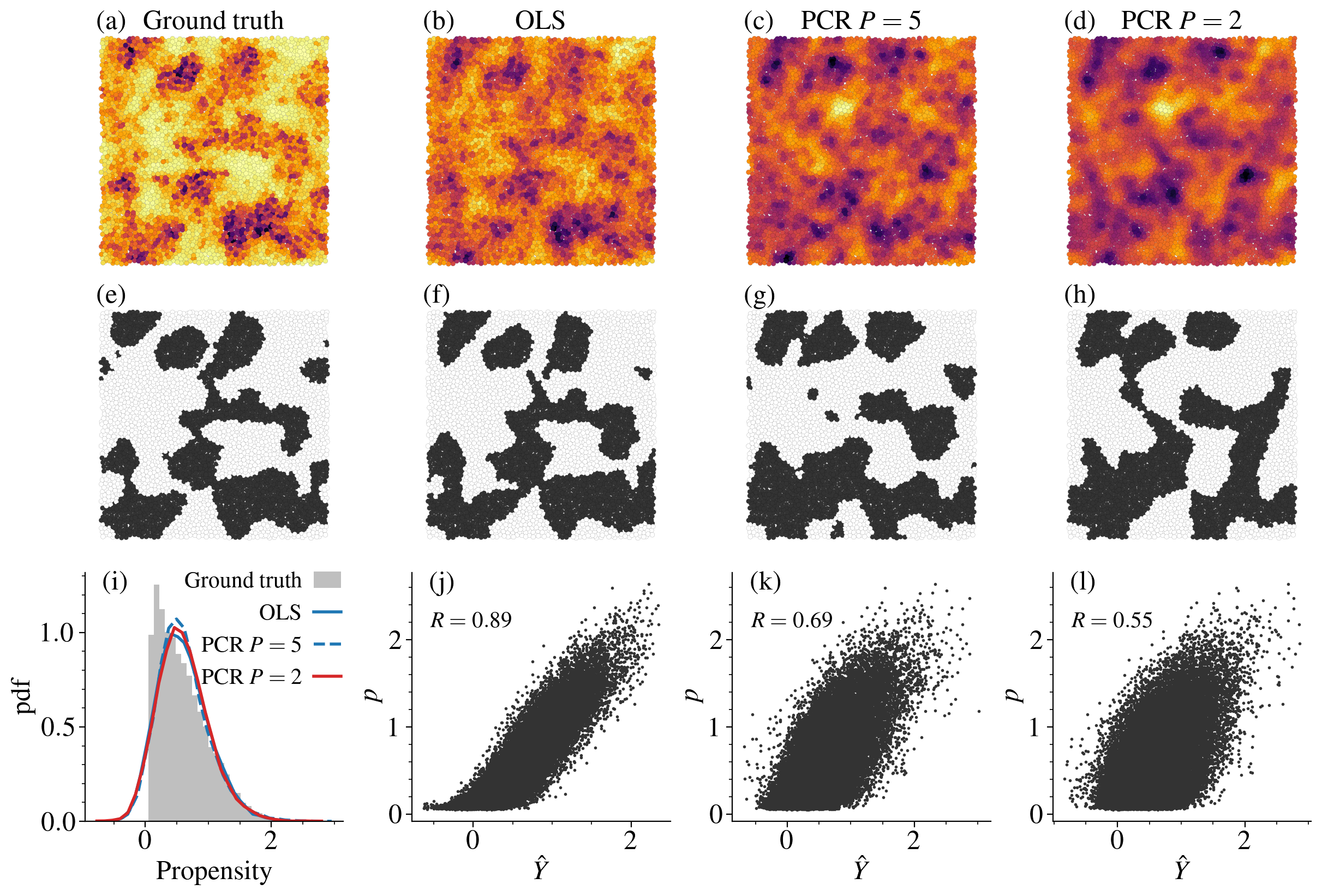}
\caption{Dynamic propensity of a representative configuration. (a) Ground truth and estimates from (b) the OLS regression model, (c) the PCR model with $P=5$ components, and (d) the PCR model with $P=2$ components. Dark and light particles correspond to fast and slow particles, respectively. The mid panels, from (e) to (h) show the same as in the top panels coarse-grained over a length $\ell=1.5$. Particles are colored in black and white according to whether the corresponding coarse-grained variable is above or below the average, respectively. (i) Probability density function of the ground truth and estimated dynamic propensities for various models, using all the available configurations. The remaining panels in the bottom row display scatter plots of the dynamic propensity against its estimated value for (j) the OLS regression model, (k) the PCR model with $P=5$ components, and (l) the PCR model with $P=2$ components. In all the panels, $\hat{Y}$ is shifted and scaled to match the mean and variance of the dynamic propensity $p$.}
\label{fig:snapshots}
\end{figure*}

\section{Discussion}
\label{sec:discussion}

Given the range of linear models that reproduce the dynamic propensity with comparable performance accuracy, a few questions arise naturally: 
is it possible to identify an ``optimal'' model that strikes a balance between prediction accuracy and physical interpretation?
Does the choice of the descriptor play an important role?
In this section, we will provide elements to try to address these questions.

\textit{Real space structure of the models' prediction}-- A first piece of information comes from direct visual comparison of the ground truth dynamic propensities and the predictions of the linear regression models.
Here, we consider two low-dimensional PCR models, namely the ones with $P=2$ and $P=5$ within supervised PCR, as well as with the full $P=M$ model (corresponding to the OLS regression model) obtained with the BP descriptor.
In this section, we revert our initial feature normalization (see Sec.~\ref{sec:dataset}): we scale and shift the predicted propensities, $\hat{Y}_i$ so that their mean and variance match those of the actual unscaled dynamic propensity.
Thus, we will consider $\hat{Y}_i$ vs. $p_i$ in the original LJ units.

Figure~\ref{fig:snapshots} shows the distributions of $\hat{Y}_i$ vs $p_i$, the scatter plots of $\hat{Y}_i$ vs $p_i$ and the corresponding snapshots for a representative configuration in our dataset.
We see that the full descriptor ($P=M$ or OLS regression model) provides an excellent description of the dynamic propensity field, which is accurately reconstructed in most of its details, except for a slight discrepancy in the shape of the distribution, see Fig.~\ref{fig:snapshots}(i).
Thus, $R$ values of the order of 0.9 correspond to a very satisfactory description of the dynamic propensity.
The model obtained with the $P=5$ most correlated PCs gives $R\approx 0.7$ and reproduces most of the patterns of the dynamic propensity field, despite some additional noise.
The correspondence is good when considering coarse-grained fields (see the mid panels), indicating that such models are able to grasp the relevant dynamic fluctuations on large length scales~\cite{berthier2007structure}.
The model with the $P=2$ most correlated PCs, instead, only captures some of the fast and slow regions of the propensity, but the overall large-scale structure of the propensity field is not accurately reproduced.
Qualitatively, these results suggest that correlation coefficients of 0.55, 0.7, 0.9 correspond to modest, satisfactory, and excellent estimation of the dynamic propensity, respectively.

\textit{Dependence on the descriptor}-- How sensitive are the results to the choice of the descriptor?
From Table~\ref{table:performance} we see that the prediction performance of the optimal Ridge regression model is rather insensitive to it. 
Note that this similarity is partly due to using the inherent structure configurations: when the calculations are performed on the instantaneous configurations, the prediction performance improves slightly with increasing the dimensionality of the descriptor (not shown).

Even when using inherent structures, however, a clear difference can be seen at the level of low-dimensional models: the low-dimensional models obtained from PCR using the SLO descriptor perform better than those obtained using the BP and JBB descriptors.
This might be explained by the inclusion from the outset of some physically relevant features like the $\Theta$ parameter.
With the SLO descriptor, in fact, even a two-features model achieves a high Pearson coefficient ($R\approx 0.81$) and very little is gained by including the full set of features ($R \approx 0.86$).
By contrast, high-dimensional descriptors accurately reconstruct the observable of interest, exploiting by brute force a broad spectrum of features, 
but they may fail to provide a simple parsimonious model.
This could also partly depend on the details of the descriptor: the atomic cluster expansion, for instance, successfully predicts plasticity in amorphous solids even upon strong dimensional reduction~\cite{rottler2024analysis}.

\begin{figure}
\includegraphics[width=\linewidth]{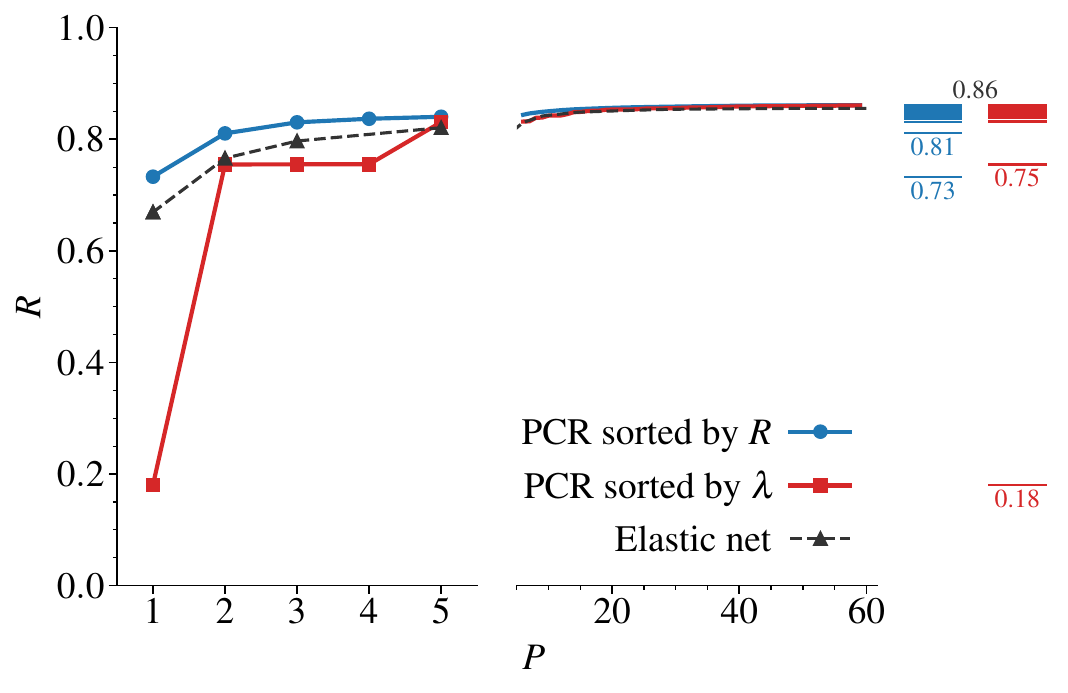}  
\caption{Same as Fig.~\ref{fig:pcr} but for the SLO descriptor.}
\label{fig:pcr_sharma}
\end{figure}

\textit{Interpretation of the structural modes}-- The above findings call for an inspection of the dominant modes in PCR using a physically motivated descriptor.
In Fig.~\ref{fig:pcr_sharma}, we show the PCR results for the SLO descriptor.
Again, the PCs that are most correlated with the dynamic propensity are not necessarily the ones with the largest eigenvalues.
We found that the modes with the largest correlations are PC2 and PC5, whose Pearson coefficients with the dynamic propensity sum up geometrically to 0.81.

\begin{figure}
\includegraphics[width=\linewidth]{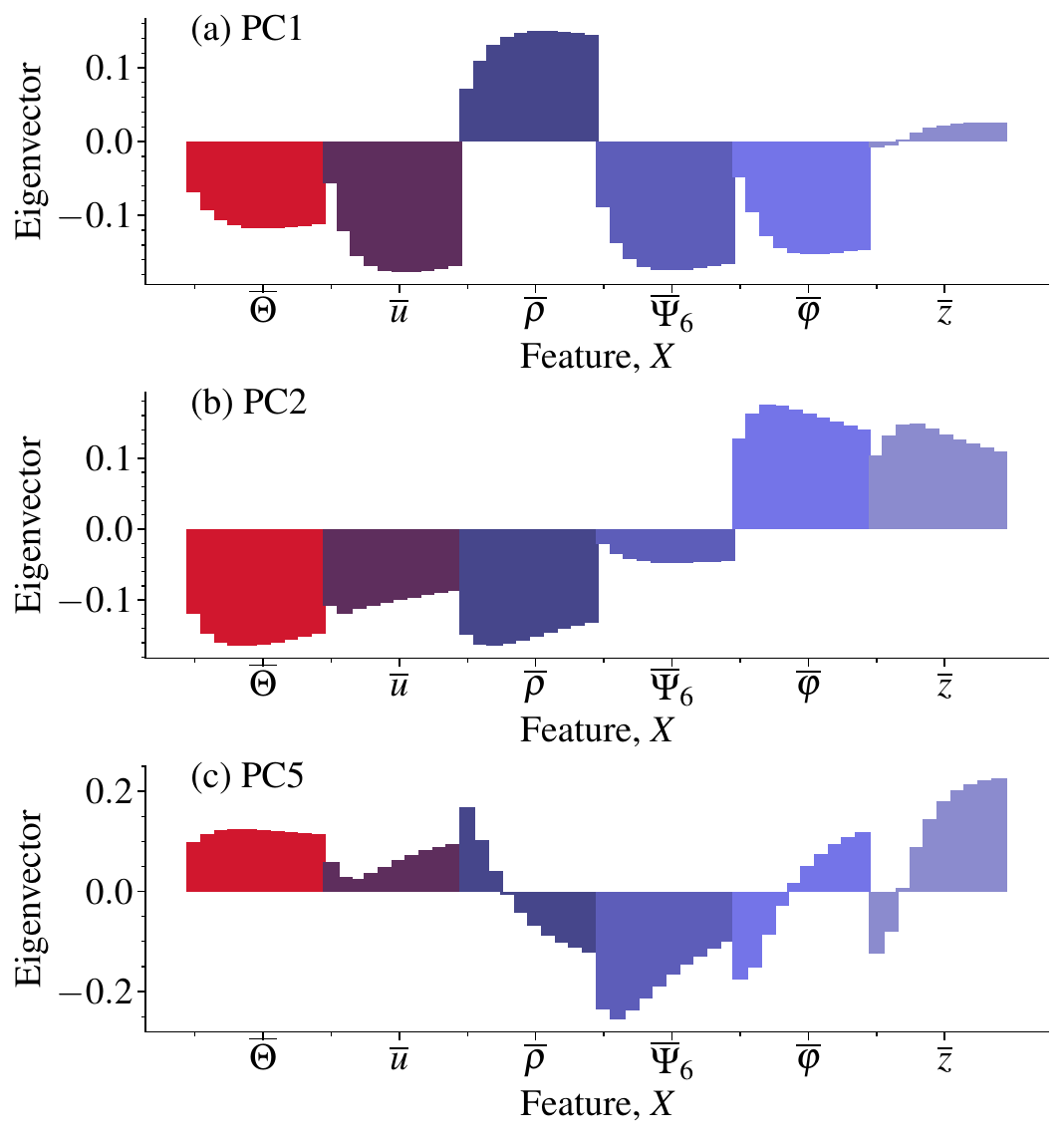}  
\caption{Eigenvectors $\mathbf{u}^{(1)}$, $\mathbf{u}^{(2)}$, and $\mathbf{u}^{(5)}$ obtained using the SLO descriptor.}
\label{fig:eigenvectors_sharma}
\end{figure}

In Fig.~\ref{fig:eigenvectors_sharma}, we show the eigenvectors corresponding to PC1, PC2, and PC5.
Let us first focus on the PC that is most relevant for the dynamics.
PC2 account for fluctuations of $\overline \Theta$ that are anti-correlated to the local packing $\overline \varphi$ (and correlated to $\overline \rho$, as expected).
This makes sense, as small values of $\overline \Theta$ should capture sterically favored environments with high local packing.
Note that the coarse-graining length associated with the larger contribution on this PC is around $\ell \approx 2.5$, which roughly corresponds to the second coordination shell.
We thus identify PC2 with a structural mode associated with the fluctuations of local packing on an intermediate length scale.

Interestingly, PCA reveals that some of the fluctuations of $\overline \Theta$ are positively correlated to $\overline \varphi$.
This unexpected behavior is described by PC1, which captures the largest fraction of the variance of the normalized structural dataset.
Our supervised PCR scheme effectively removes this source of fluctuations, which is irrelevant for the dynamics ($R=0.18$).
Note that the fluctuations described by PC2 are more strongly correlated to the dynamic propensity than the bare $\overline \Theta$ parameter.

Turning our attention to PC5, we see that the largest contribution to this mode comes from the 6-fold orientational order parameter, $\overline \Psi_6$, averaged over a short range distance ($\ell \approx 1.0$).
We point out, however, that the relevant bond-orientational order of our ternary model need not be 6-fold.
Indeed, previous analysis of the low-energy states of the binary version of the model indicated the presence of quasi-crystalline order with 5- and 10-fold symmetries.
We leave a more systematic analysis of the preferred bond-orientational order of our ternary model for a future study.

To sum up on this point, PCR of the SLO descriptor yields a simple two-feature structural model.
The two relevant variables account for steric effects on an intermediate length scale and for some of the fluctuations of bond-orientational order.
Superficially, these results suggest a connection with the phenomenological two-state model developed by Tanaka in the 1990s~\cite{tanaka1998simple}, which included local density and bond-order fluctuations as structural order parameters.
We emphasize, however, that our ternary model does not display strong 6-fold orientational order and $\overline{\Psi}_6$ itself is poorly correlated to the dynamic propensity.
It is only when the fluctuations of $\overline{\Psi}_6$ are coupled to those of the other features as described by $\textbf{u}^{(5)}$ that some correlation emerges.
We found that the other descriptors consistently identify a structural mode related to local packing ({\it e.g.}, PC2 in the BP descriptor, PC1 in the JBB descriptor) that is moderately correlated to the dynamics, but none that is specifically connected to bond-orientational order.

\begin{figure}
\includegraphics[width=\linewidth]{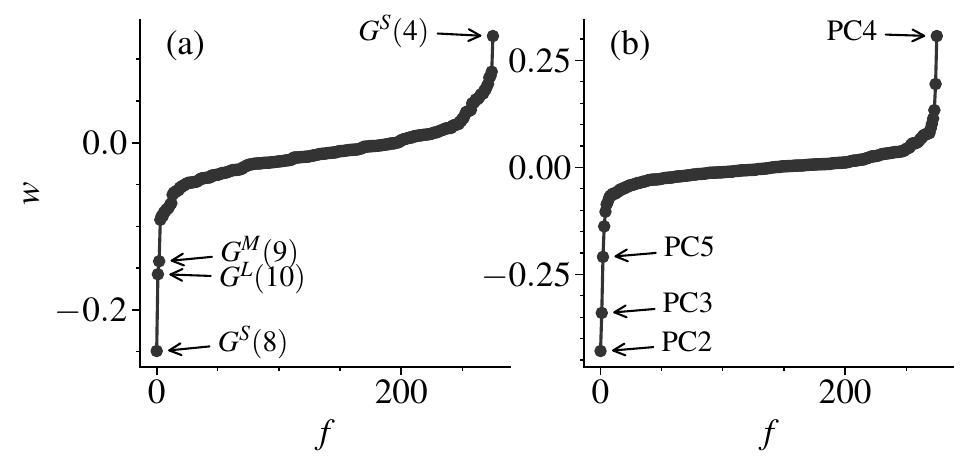}  
\caption{Sequence of sorted weights $w$ as a function of sorted feature index $f$ in (a) Ridge regression with $\alpha=0.1$ and (b) PCR. Note that in PCR the weights coincide with the Pearson coefficients $\R{\tilde{X}}{Y}$, see Eq.~\eqref{eq:Y_PCA_OLS}.}
\label{fig:gaps}
\end{figure}

\textit{Feature selection}-- One practical approach to feature selection is to look for gaps in the weight distribution of the linear models: features whose absolute weights stand out from the bulk are then identified as the most relevant ones.
In Fig.~\ref{fig:gaps}, we illustrate these ideas for Ridge regression with $\alpha=0.1$ and PCR using the BP descriptor.
In each panel, we show the sequence of sorted weights $w$ as a function of sorted feature index $f$.
By visual inspection, we see that no more than four or five features show weights that stand out from the rest in such an ordered sequence.
A preliminary analysis of the weight distributions support this idea.
In Ridge regression, the relevant features pinpoint structural modes involving radial features localized around the first peaks of the partial radial distribution functions.
In PCR, the principal components from PC2 to PC5 stand out, as was also clear from Fig.~\ref{fig:pca}.
Interestingly, we found that the two features with largest absolute weights in Ridge regression, \textit{i.e.}, $G^{S}(8)$ and $G^{L}(10)$, have similar projections on the physically motivated variables as PC1 and PC2, respectively.
These weights' sequences provide good examples of sparse representations that support physical interpretation.

\textit{Optimality criteria}-- Are there quantitative criteria to decide whether to accept a low-dimensional model, which is easier to interpret, at the expense of performance?
We considered several optimality criteria, developed in the context of statistical learning, that attempt to define an optimum model by balancing the performance and the model complexity.
In particular, we employed two well-known scores: the Akaike information criterion~\cite{akaike2003new} and the Bayesian information criterion~\cite{schwarz1978estimating}. 
Both criteria address over-fitting by combining the maximized log-likelihood, which quantifies data fidelity, with a complexity penalty term that grows with the number of parameters.
We found, however, that these criteria do not provide any practical guidance in our setting, since we operate in a high-data regime where overfitting is negligible.

Our current standpoint on this issue is that at present there is no robust quantitative criterion to select an optimal model that strikes a balance between prediction and model size.
Selecting a small number of physical variables, say up to five, is in line with a well-established approach to develop phenomenological models in statistical physics~\cite{kivelson2018understanding}.
We thus suggest that a parsimonious data-driven description of glassy dynamics should involve no more than a handful of relevant features.

\textit{Dependence on spatial dimensions}-- Characterizing the local structure of glassy systems may appear an easier task in 2D than in 3D.
Provided the input structural descriptor is complete enough, however, linear regression and deep learning models of glassy dynamics eventually achieve similar performances in both spatial dimensions~\cite{jung2025roadmap}.
While 3D systems may pose a more serious challenge for dimensional reduction, the same kind of physically motivated features we used in this study as explanatory variables can be readily applied to 3D systems.
Therefore, at least from the methodological viewpoint, there is no qualitative difference between studying 2D and 3D systems.

\begin{figure}
\includegraphics[width=\linewidth]{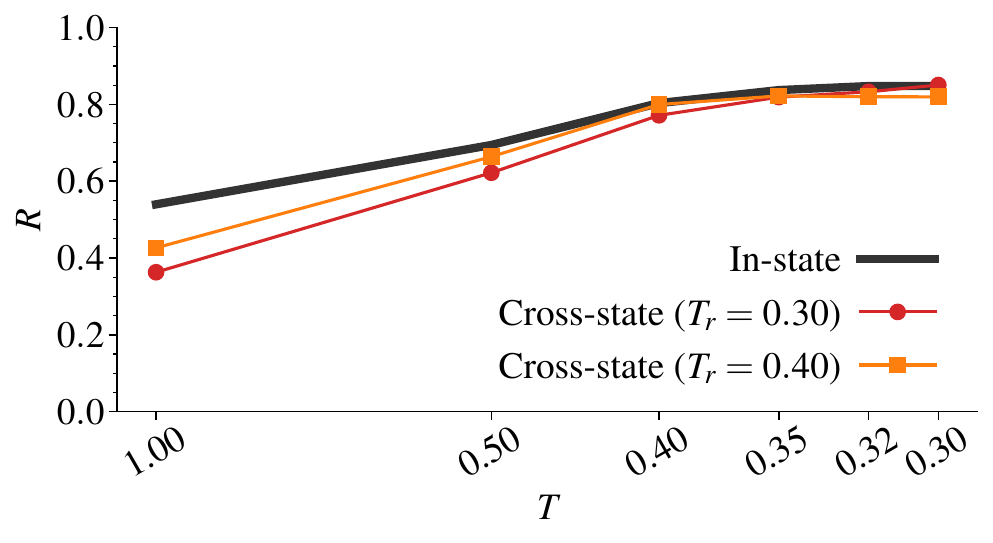}  
\caption{Cross-state generalization of the propensity of motion $p(t=\tau_\alpha; T)$ from Ridge regression at a reference temperature $T_r$. The solid line indicates the in-state prediction (interpolation), while the thin lines with symbols indicate the cross-state predictions (extrapolation), from $T_r=0.30$ (circles) and $T_r=0.40$ (squares).}
\label{fig:prediction}
\end{figure}

\textit{Cross-state generalization}-- Throughout this work we focused on the performance of linear regression at the same temperature at which the model was trained. As anticipated in Sec.~\ref{sec:problem}, from a physicist's viewpoint, data interpolation does not qualify as a genuine form of prediction. To assess the actual predictive power of Ridge regression, we show in Fig.~\ref{fig:prediction} the results for the cross-state generalization of the propensity of motion at $t=\tau_\alpha$: the model's weights are fitted at a reference temperature $T_r$ using $\alpha=0.1$ and then used to extrapolate the propensity of motion at a different temperature $T$. The extrapolation from $T_r=0.30$ to higher temperatures provides good predictions up to around the onset temperature $T\approx 0.5$, where the Pearson coefficients start to decrease significantly even for in-state predictions. Interestingly, the model trained at $T_r=0.40$ provides consistent predictions over a broad range of temperatures, covering around three decades in structural relaxation times~\cite{sharma2024selecting}. These findings resemble those of obtained in Ref.~\onlinecite{jung2025roadmap} using linear regression and deep learning models. Note, however, that all these ML models can only predict the propensity of motion at a time $t=\tau_\alpha$, \textit{i.e.}, the absolute time scale must be determined independently. Thus, current ML models are not fully predictive yet.

\section{Conclusions}
\label{sec:conclusions}

Recent work has shown that data-driven models, based on either deep neural networks or linear regression, can accurately describe and predict the dynamic propensity of glass-forming liquids on the structural relaxation time scale~\cite{jung2025roadmap}.
In our opinion, the current level of prediction accuracy of these models is high enough to motivate a shift of focus toward interpretation~\cite{swain2024machine}.
In fact, a common criticism to these machine learning studies is that they still provide little physical insight into the underlying mechanisms behind glassy dynamics.
Identifying interpretable models that provide a robust and succinct relationship between physically relevant variables is crucial to address this issue.

The main goal of this paper was to assess and improve the interpretability of linear regression models of glassy dynamics using a simple model of two-dimensional glass.
We showed that, contrary to previous expectations~\cite{jung2025roadmap}, even linear regression models can be hard to interpret.
A major issue can be traced back to the presence of strong linear dependencies between structural features, known as multicollinearity, which hinders the interpretability of linear models.
We found that several structural descriptors used in recent studies are severely affected by multicollinearity.
Ridge regression can be used to suppress some of the detrimental effects of multicollinearity, but the resulting models are not succinct enough to be interpretable. 

To identify low-dimensional linear models of glassy dynamics, we used two simple dimensional reduction techniques.
First, we considered elastic net regression, which yields a set of low-dimensional models that strike a good balance between accuracy and physical interpretability.
Second, we performed principal component regression using a supervised selection scheme of the principal components.
This approach yields an accurate enough description of the dynamic propensity with a handful of collective structural features. 
Overall, our work establishes that linear regression models, once properly fine-tuned, can be useful tools to identify the relevant structural modes associated to dynamic heterogeneities in glass-forming liquids.
It would be interesting to extend our analysis to more sophisticated high-dimensional descriptors and to other glassy systems beyond the simple two-dimensional model liquid considered here, including molecular and polymeric liquids as well as amorphous solids subject to external loading.

\begin{acknowledgments}
We thank Gerhard Jung for insightful discussions.
MO thanks the support by MIAI@Grenoble Alpes and the Agence Nationale de la Recherche under France 2030 with the reference ANR-23-IACL-0006. 
\end{acknowledgments}

\section*{Data availability}
The data and workflow necessary to reproduce the findings of this study are available in the Zenodo data repository~\cite{zenodo}. 

\appendix

\section{Species-resolved analysis}
\label{sec:appendix_species_dependence}

In multi-component mixtures, the particle mobility is in general species dependent.
Therefore, most studies of structure-dynamics relationship analyze separately each species, or even just focus on the majority species.
In the body of the text, we chose to analyze all the particles irrespective of their species.
As we will show in this appendix, the key findings of our work do not depend on whether the analysis is performed species-wise or for the full system.

\begin{figure}
\includegraphics[width=\columnwidth]{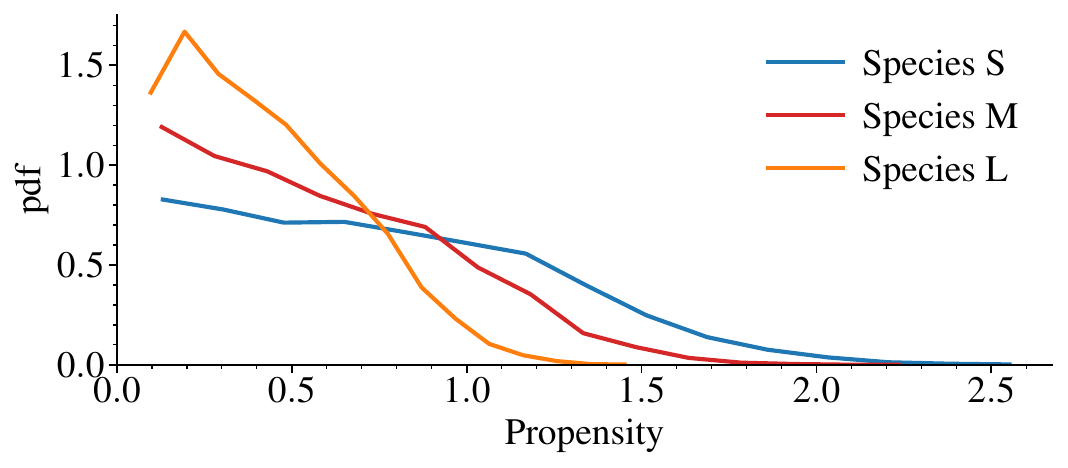}  
\caption{Probability density function of the dynamic propensity for particles of species S (blue), M (red), and L (orange).}
\label{fig:species_propensity}
\end{figure}

In Fig.~\ref{fig:species_propensity} we show the probability density functions for the dynamic propensity for particles of species S, M, and L.
The distribution functions depend of course on the species, although the dependence is not dramatic.
We found that the average values and corresponding standard deviations of the distributions are $0.75 \pm 0.47$, $0.58 \pm 0.37$, and $0.43 \pm 0.25$ for species S, M, and L, respectively.
For comparison, the corresponding result for the dynamic propensity of all the particles is $0.60 \pm 0.41$.
Since we could not find detailed results regarding species-dependent propensities in the literature~\cite{jung2024dynamic}, we cannot determine whether the above differences are smaller than those for other 3d glass-forming liquids.

\begin{figure}
\includegraphics[width=\columnwidth]{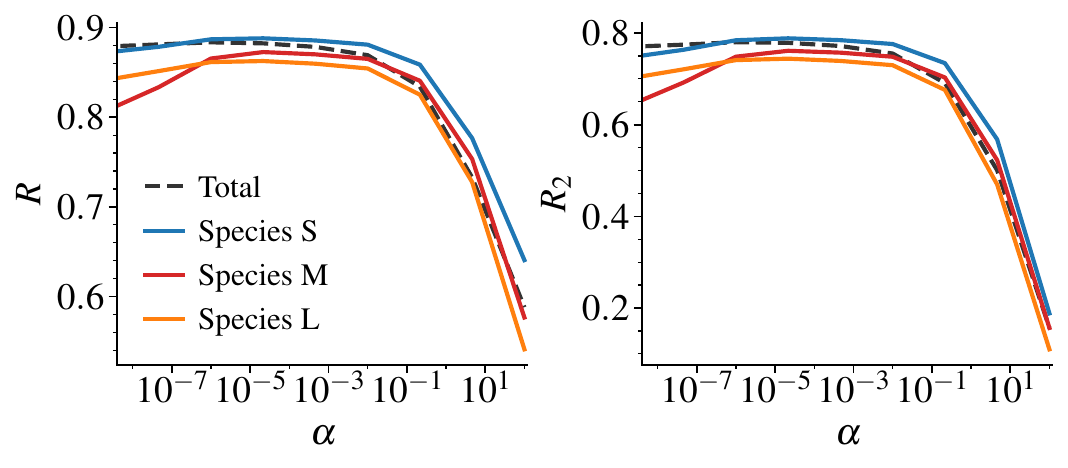}  
\caption{Species-wise (full lines) and total (dashed lines) prediction performance metrics for Ridge regression of the dynamic propensity using the BP descriptor: (a) Pearson coefficient $R[Y, \hat{Y}]$ as a function of $\alpha$ and (b) coefficient of determination $R_2[Y, \hat{Y}]$ as a function of $\alpha$.}
\label{fig:species_ridge}
\end{figure}

One might expect that analyzing structure-dynamics correlations for the entire system will yield a different performance compared to a species-wise analysis.
As shown in Fig.~\ref{fig:species_ridge}, this is indeed the case, but the effect is very small.
In particular, for the optimal regularization parameter $\alpha=0.1$, we obtained Pearson coefficients between the BP descriptor and the propensity of
0.86, 0.85, 0.84 for species S, M, and L, respectively.
These values can be compared with $R=0.85$ obtained for the full system, see Table.~\ref{table:performance}.
Somewhat larger discrepancies are observed at very small $\alpha$, or in the absence of regularization, for the minority species (M).
Overall, however, the performance of Ridge regression displays the same qualitative trends when performed species-wise or for the full system.
We also confirm the presence of oscillations in the weights for $\alpha=0$ in the species-wise analysis.

\begin{figure}
\includegraphics[width=\columnwidth]{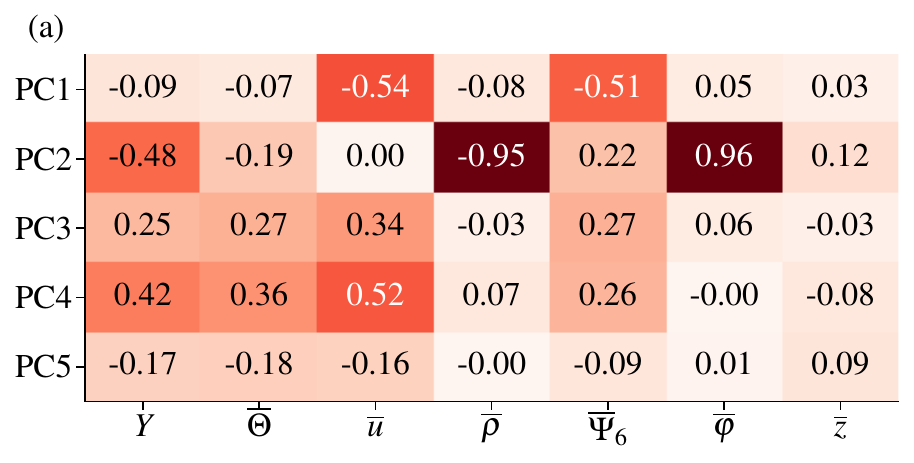}  
\includegraphics[width=\columnwidth]{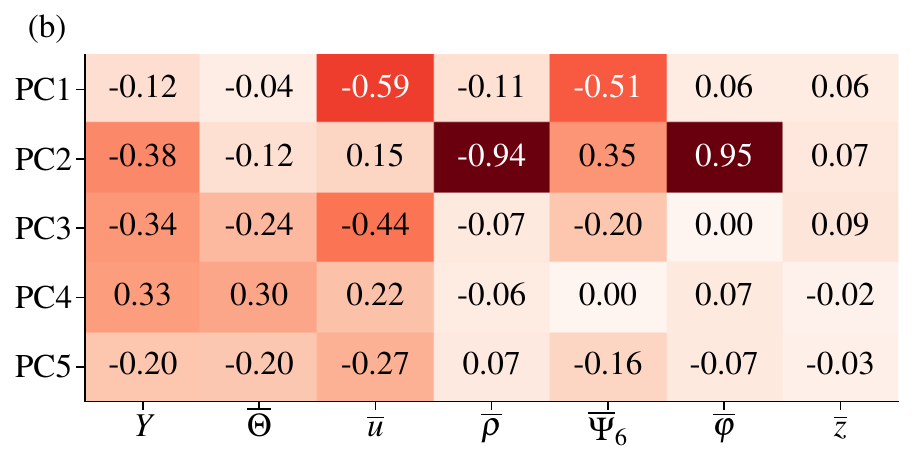}  
\includegraphics[width=\columnwidth]{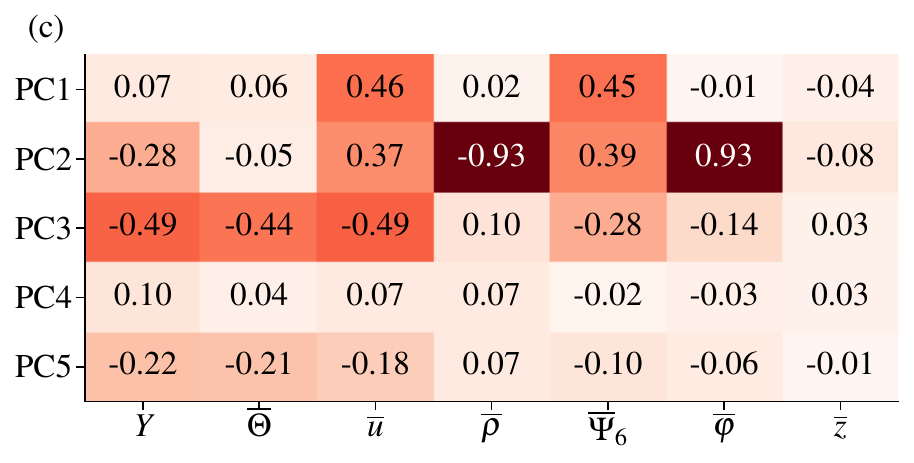}  
\caption{Pearson coefficients $R$ between the first five PCs and physically motivated features coarse-grained over a length $\ell=1.5$. The analysis is performed species-wise for (a) species S, (b) species M, and (c) species L.}
\label{fig:species_eigenvectors}
\end{figure}

We now turn our attention to the PCA of the BP descriptor.
Overall, the trend of PCA is qualitatively similar when performed species-wise or for the full system.
The observation that PC1 is poorly correlated to the dynamic propensity is confirmed for all species.
We found that PC2 is again the most correlated for species S and M, as for the full system. For species L, it is PC3 to be most correlated.
These results suggest that it may be difficult to extract robust information from PCA beyond the very first few PCs (in $\lambda$ order).

Interestingly, we found that also the physical interpretation of the first two PCs is essentially the same across species.
In Fig.~\ref{fig:species_eigenvectors}, we show at the Pearson coefficients between the first 5 PCs of the BP descriptor and the physically motivated features.
The latter features are coarse-grained over $\ell=1.5$ and the analysis is performed species-wise in each panel.
The key structure of the correlation matrix for the species-resolved analysis is similar to the one of full system.
In fact, PC1 and PC2 are projected similarly on the physically resolved features.
Note, however, that for species L it is PC3 having the largest correlation with the dynamics and not PC2.
It is difficult to grasp a clear interpretation of components of order higher than two in terms of physically motivated features.

\begin{figure}
\includegraphics[width=\columnwidth]{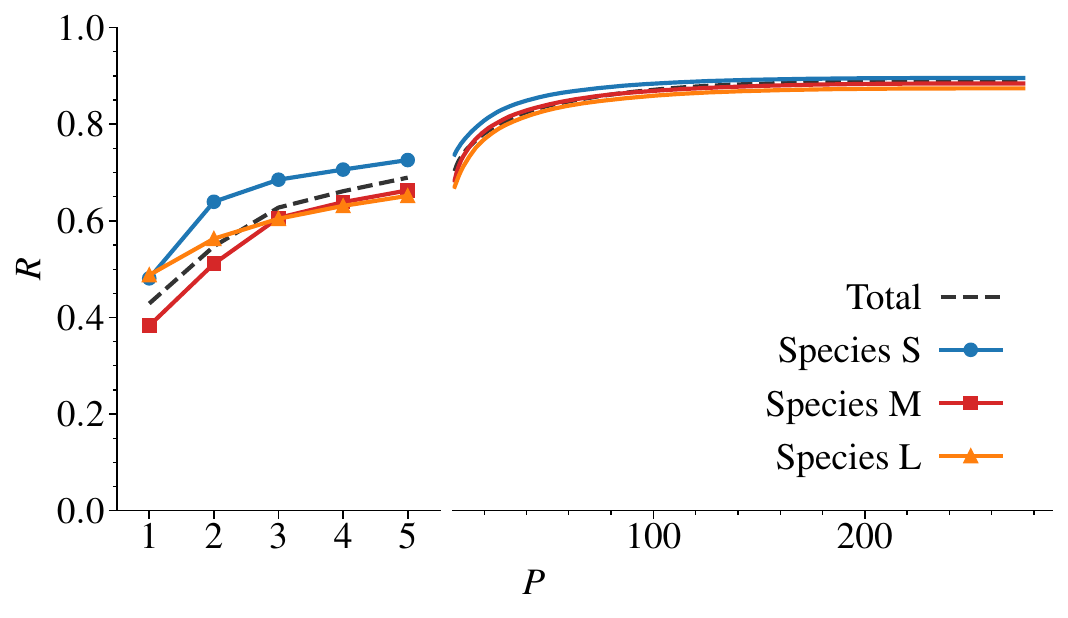}  
\caption{Species-wise (full lines) and total (dashed lines) Pearson coefficient $R$ between the ground truth dynamic propensity and the PCR predictions using the first $P$ PCs sorted according to their eigenvalue (squares). Note the change of scale on the x-axis after $P=5$ components.}
\label{fig:species_pcr}
\end{figure}

Finally, we assess the species dependence of the performance of PCR.
Figure~\ref{fig:species_pcr} shows the Pearson coefficient $R$ in PCR as function of the number $P$ of included collective features.
The latter are sorted according to their individual Pearson coefficients.
The growth of the Pearson coefficient as a function of $P$ is qualitatively similar for all the species and closely tracks the one for the total system.
We conclude that although the details about the PCs with small eigenvalues and small correlation will depend in general of the selected species, the general trends discussed in the main text are robust.

\section{Selected results for the SLO and JBB descriptors}
\label{sec:appendix_datasets}

\begin{figure}
\includegraphics[width=\columnwidth]{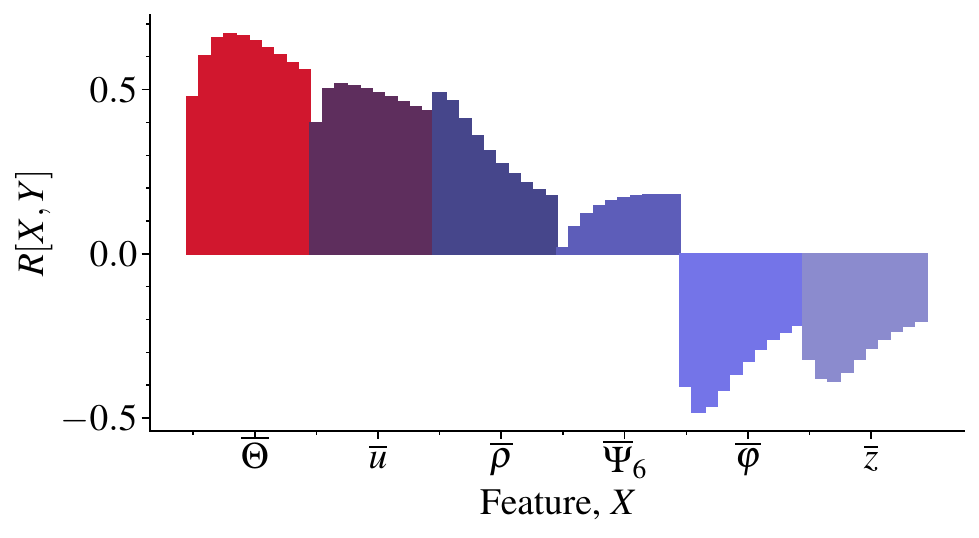}  
\caption{ 
Pearson coefficient, $\R{X^{(f)}}{Y}$, between the dynamic propensity $\textbf{Y}$ and each structural feature $\textbf{X}^{(f)}$ of the SLO descriptor for $f=1, \dots, M$.}
\label{fig:pearson_slo}
\vskip1em
\includegraphics[width=\columnwidth]{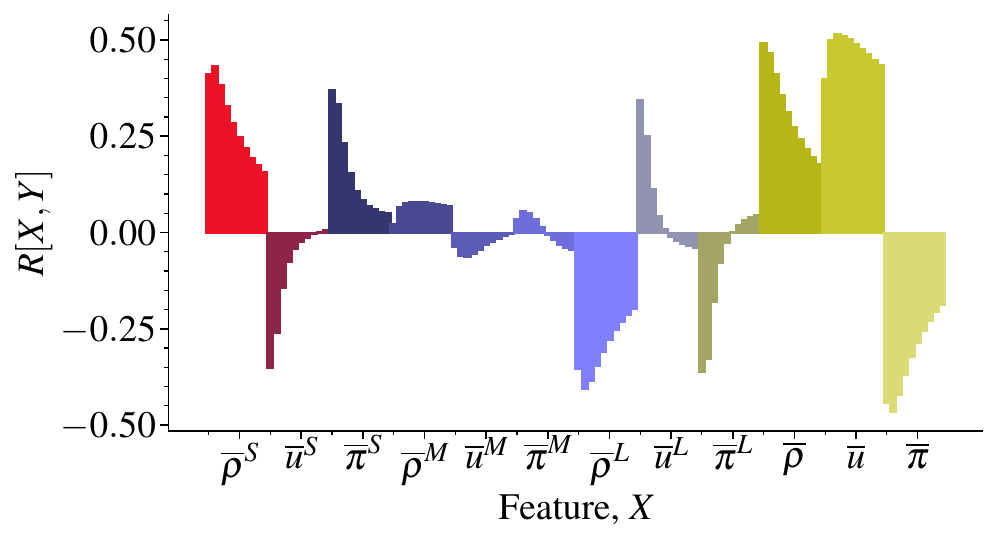}  
\caption{ 
Pearson coefficient, $\R{X^{(f)}}{Y}$, between the dynamic propensity $\textbf{Y}$ and each structural feature $\textbf{X}^{(f)}$ of the JBB descriptor for $f=1, \dots, M$.}
\label{fig:pearson_jbb}
\end{figure}

In this appendix, we present additional results for the physically motivated descriptors introduced in Sec.~\ref{sec:physical_descriptors}.
Figures~\ref{fig:pearson_slo} and \ref{fig:pearson_jbb} display the Pearson coefficients $\R{X^{(f)}}{Y}$ between the dynamic propensity and each structural feature $\textbf{X}^{(f)}$ of the SLO and JBB descriptors, respectively.

The SLO descriptor emphasizes the substantial role of packing efficiency in determining dynamic fluctuations.
This can be appreciated by the large positive correlations between $\overline{\Theta}$ and the dynamic propensity, as well as by the negative correlations with $\overline{\varphi}$.
Note, however, that the coarse-graining length $\ell$ that gives the largest correlation is not the same for $\overline{\Theta}$ and $\overline{\varphi}$.
The bond-orientational order parameter, $\overline{\Psi}_6$, by contrast, is poorly correlated with the dynamics, see the discussion in Sec.~\ref{sec:pcr}.

It is instead more difficult to extract straightforward physical information from the analysis of the JBB descriptor, owing to its complex dependence on the chemical composition.
It is nonetheless interesting to note that the perimeter $\overline \pi$ of the Voronoi cells, which is included in the JBB descriptor but not in the SLO one, has an overall negative correlation with the dynamic propensity.
This again points to a coupling between steric effects, \textit{i.e.}, larger local free volume, and mobility.

\begin{figure}
\includegraphics[width=\columnwidth]{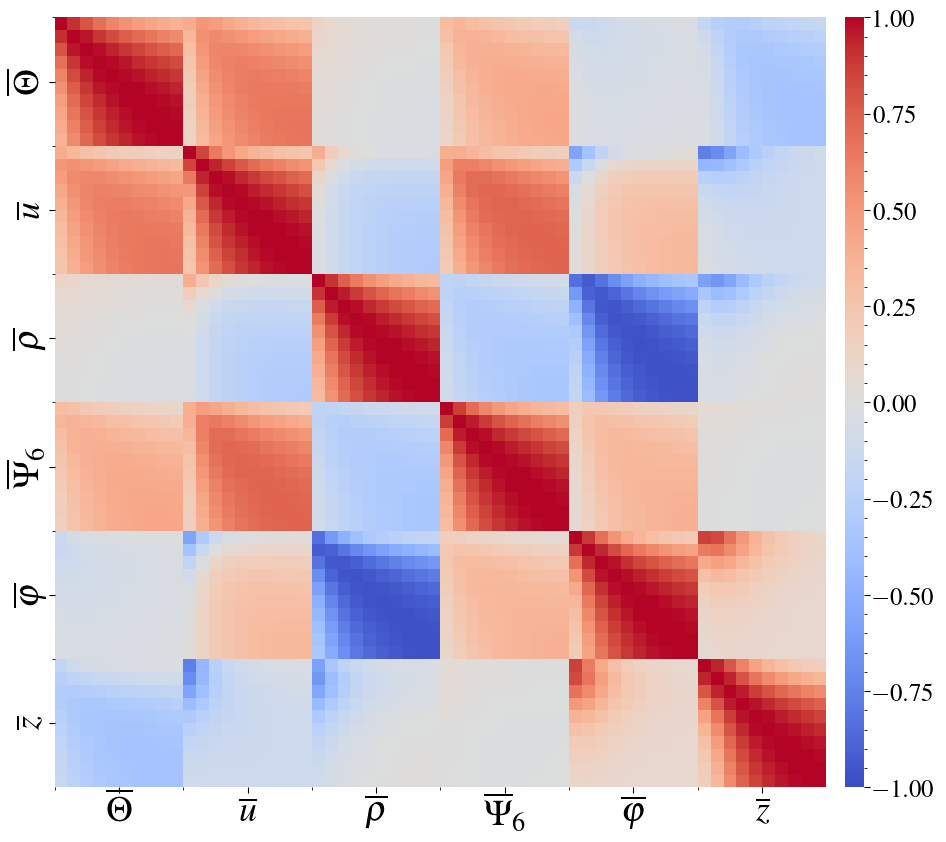}  
\caption{ 
Correlation matrix $\mathrm{C}$ for the SLO descriptor.}
\label{fig:correlation_matrix_slo}
\vskip1em
\includegraphics[width=\columnwidth]{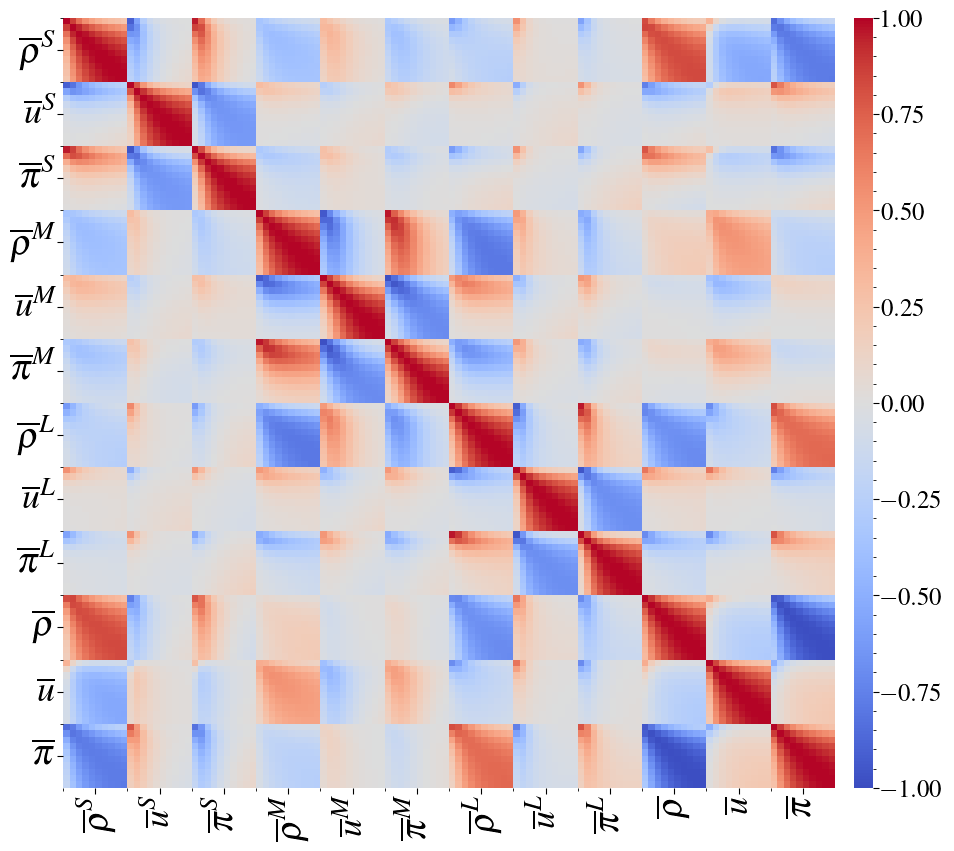}  
\caption{ 
Correlation matrix $\mathrm{C}$ for the JBB descriptor.}
\label{fig:correlation_matrix_jbb}
\end{figure}

In Figures~\ref{fig:correlation_matrix_slo} and~\ref{fig:correlation_matrix_jbb}, we show  the correlation matrices $\mathrm{C}$ of the SLO and JBB descriptors, respectively.
As for the BP descriptor, the correlation matrices display a pronounced block structure: there are strong correlations within subblocks of features, this time due to similar coarse-graining lengths, as well as non-trivial correlations between different features.
In particular, the SLO descriptor reveals a negative correlation between local number density $\overline{\rho}$ and local packing fraction $\overline{\varphi}$.
This counter-intuitive result is likley a result of compositional fluctuations in this mixture, see Sec.~\ref{sec:pearson}.

\section{Ordinary least squares regression and Ridge regression}
\label{sec:appendix_regression}

We derive the estimation of the weights $\hat{\bf w}$ by the ordinary least squares regression (OLS) and Ridge regression for the dataset, $(\mathrm{X}, {\bf Y})$, defined in Eqs.~(\ref{eq:Y_vector}) and (\ref{eq:design_matrix}).

We first consider a linear model,
\begin{equation}
    \hat{\bf Y} = \mathrm{X} \hat{\bf w},
\end{equation}
where $\hat{\bf Y} = [\hat{Y}_1, \hat{Y}_2, \ldots, \hat{Y}_{N_\mathcal{S}}]^T$ represents the predictions and $\hat{\bf w} = [\hat{w}^{(1)}, \hat{w}^{(2)}, \ldots, \hat{w}^{(M)}]^T$ denotes the weight parameters.

The loss function $\mathcal{L}^{\rm Ridge}(\hat{\bf w})$ in Eq.~\eqref{eq:loss_ridge} can be rewritten as
\begin{align}
    \mathcal{L}^{\rm Ridge}(\hat{\bf w}) &= \frac{1}{2 N_\mathcal{S}} \left( \hat{\bf w}^T \mathrm{X}^T \mathrm{X} \hat{\bf w} - 2 \left(\mathrm{X}^T {\bf Y}\right)^T \hat{\bf w} + {\bf Y}^T {\bf Y} \right) \nonumber \\
    &\quad + \frac{\alpha}{2} \hat{\bf w}^T \hat{\bf w}.
\end{align}
The derivative of $\mathcal{L}^{\rm Ridge}(\hat{\bf w})$ with respect to $\hat{\bf w}$ is
\begin{equation}
    \nabla_{\hat{\bf w}} \mathcal{L}^{\rm Ridge}(\hat{\bf w}) = \left(\frac{1}{N_\mathcal{S}} \mathrm{X}^T \mathrm{X} + \alpha \mathrm{I}\right) \hat{\bf w} - \frac{1}{N_\mathcal{S}} \mathrm{X}^T {\bf Y},
    \label{eq:derivative_loss}
\end{equation}
where $\mathrm{I}$ is the $M \times M$ identity matrix.

Since $Y_i$ (elements of ${\bf Y}$) and $X_i^{(f)}$ (elements of $\mathrm{X}$) are normalized to have zero mean and unit variance, we can express the terms in Eq.~(\ref{eq:derivative_loss}) using the correlation matrix $\mathrm{C}$ defined in Eq.~(\ref{eq:correlation_matrix}) and the Pearson coefficients $R$,
\begin{eqnarray}
    \frac{1}{N_\mathcal{S}} \mathrm{X}^T \mathrm{X} &=& \mathrm{C} , \\
     \frac{1}{N_\mathcal{S}} \mathrm{X}^T {\bf Y} &=& \left[\R{X^{(1)}}{Y}, \ \ldots, \ \R{X^{(M)}}{Y} \right]^T \nonumber \\ 
     &=& \Rb{X}{Y} .  
\end{eqnarray}

Setting $\nabla_{\hat{\bf w}} \mathcal{L}^{\rm Ridge}(\hat{\bf w}) = {\bf 0}$, the solution for Ridge regression is obtained as
\begin{equation}
    \hat{\bf w}_{\rm Ridge} = (\mathrm{C} + \alpha \mathrm{I})^{-1} \Rb{X}{Y}.
\end{equation}

\section{Condition number of a matrix}
\label{sec:appendix_condition_number}

We briefly review the condition number of a matrix, without going into the details of a mathematically rigorous treatment. It is defined using the norm of a matrix, and we also review its meaning as a measure of a matrix's instability. One can see detailed discussions in, {\it e.g.}, Refs.~\onlinecite{golub2013matrix,trefethen2022numerical}.

\subsection{Matrix norm}

A matrix norm is a generalization of the absolute value $|\cdot|$ for scalars and the vector norm $||\cdot||$ for vectors, applied to matrices. The matrix norm is also denoted as $||\cdot||$.  
Here, we summarize only the important properties relevant for our purposes. 
For a scalar $\alpha$, a vector ${\bf x}$, and matrices $\mathrm{A}$ and $\mathrm{B}$, the following properties hold:  
i) $||\mathrm{A}|| \geq 0$,  
ii) $||\alpha \mathrm{A}|| = |\alpha| \, ||\mathrm{A}||$,  
iii) $||\mathrm{A} + \mathrm{B}|| \leq ||\mathrm{A}|| + ||\mathrm{B}||$,  
iv) $||\mathrm{A}{\bf x}|| \leq ||\mathrm{A}|| \, ||{\bf x}||$,  
v) $||\mathrm{A}\mathrm{B}|| \leq ||\mathrm{A}|| \, ||\mathrm{B}||$, and  
vi) $||\mathrm{I}|| = 1$, where $\mathrm{I}$ is the identity matrix.

There are various ways to define the matrix norm that satisfy these properties. One convenient approach is to use the maximum singular value, $s_{\rm max}$, of a matrix. The norm $||\mathrm{A}|| = s_{\rm max}$ is referred to as the spectral norm.

For a positive semi-definite symmetric matrix, the singular values are equal to the eigenvalues. Thus, we have $||\mathrm{A}|| = \lambda_{\rm max}$, where $\lambda_{\rm max}$ is the largest eigenvalue.  
Moreover, if $\mathrm{A}$ is invertible, then $||\mathrm{A}^{-1}|| = \lambda_{\rm min}^{-1}$, where $\lambda_{\rm min}$ is the smallest eigenvalue.

\subsection{Definition of condition number}

In general, the condition number $\kappa(\mathrm{A})$ is defined as 
\begin{equation}
    \kappa(\mathrm{A}) = ||\mathrm{A}|| \, ||\mathrm{A}^{-1}|| .
    \label{eq:def_condition_number}
\end{equation}  
In this paper, we consider the spectral norm, and assume that $\mathrm{A}$ is a positive semi-definite symmetric matrix.  Thus, we obtain
\begin{equation}
    \kappa(\mathrm{A}) = \frac{\lambda_{\rm max}}{\lambda_{\rm min}} .
\end{equation}

\subsection{Instability of matrix}

The condition number $\kappa(\mathrm{A})$ corresponds to a degree of instability of a matrix $\mathrm{A}$. This concept can be best understood through perturbation analysis of a linear system.

Suppose we wish to solve the linear system
\begin{equation}
    \mathrm{A} {\bf x} = {\bf b} ,
    \label{eq:linear_system}
\end{equation}
where $\mathrm{A}$ is a square matrix, and ${\bf x}$ and ${\bf b}$ are vectors.  
When $\mathrm{A}$ is invertible, the solution is given by
\begin{equation}
    {\bf x} = \mathrm{A}^{-1} {\bf b} .
\end{equation}

In practice, measuring $\mathrm{A}$ and ${\bf b}$ involves errors $\delta \mathrm{A}$ and $\delta {\bf b}$ due to, for example, lack of statistics, numerical errors, etc.  
We now ask how the error in ${\bf x}$, denoted as $\delta {\bf x}$, is induced by errors in $\mathrm{A}$ and/or ${\bf b}$.

\vspace{0.5cm}
\noindent
1) When ${\bf b}$ has error $\delta {\bf b}$, the linear equation becomes
\begin{equation}
    \mathrm{A}({\bf x} + \delta {\bf x}) = {\bf b} + \delta {\bf b} ,
    \label{eq:linear_system_b}
\end{equation}
which, using Eq.~(\ref{eq:linear_system}), leads to $\mathrm{A}\delta {\bf x} = \delta {\bf b}$, so that
\begin{equation}
    \delta {\bf x} = \mathrm{A}^{-1}\delta {\bf b}.
\end{equation}
Thus, the relative error $||\delta {\bf x}||/||{\bf x}||$ is bounded as follows:
\begin{equation}
    \frac{||\delta {\bf x}||}{||{\bf x}||} = \frac{||\mathrm{A}^{-1}\delta {\bf b}||}{||{\bf x}||} \leq \frac{||\mathrm{A}^{-1}|| \, ||\delta {\bf b}||}{||{\bf x}||} .
    \label{eq:bounded}
\end{equation}
Additionally, since $||{\bf b}|| = ||\mathrm{A}{\bf x}|| \leq ||\mathrm{A}|| \, ||{\bf x}||$, we obtain the bound:
\begin{equation}
    \frac{1}{||{\bf x}||} \leq \frac{||\mathrm{A}||}{||{\bf b}||}.
\end{equation}
Thus, Eq.~(\ref{eq:bounded}) becomes
\begin{equation}
    \frac{||\delta {\bf x}||}{||{\bf x}||} \leq \frac{||\mathrm{A}|| \, ||\mathrm{A}^{-1}|| \, ||\delta {\bf b}||}{||{\bf b}||} .
\end{equation}

Using the definition of the condition number $\kappa(\mathrm{A})$ in Eq.~(\ref{eq:def_condition_number}), we conclude that
\begin{equation}
    \frac{\frac{||\delta {\bf x}||}{||{\bf x}||}}{\frac{||\delta {\bf b}||}{||{\bf b}||}} \leq \kappa(\mathrm{A}) .
    \label{eq:ineq_case1}
\end{equation}

The ratio between the relative error $||\delta {\bf x}||/||{\bf x}||$ and $||\delta {\bf b}||/||{\bf b}||$ quantifies how tiny perturbations or fluctuations in ${\bf b}$ influence the error in ${\bf x}$, which can be interpreted as a form of susceptibility.  
Equation~(\ref{eq:ineq_case1}) shows that this ratio is bounded by the condition number $\kappa(\mathrm{A})$.  
When $\kappa(\mathrm{A})$ is small, the linear system in Eq.~(\ref{eq:linear_system}) is stable. However, when $\kappa(\mathrm{A})$ is large, it becomes unstable.

\vspace{0.5cm}
\noindent
2) When $\mathrm{A}$ has error $\delta \mathrm{A}$, one can similarly derive the inequality for the relative error $||\delta {\bf x}||/||{\bf x} + \delta {\bf x}||$ when $\mathrm{A}$ has an error $\delta \mathrm{A}$:
\begin{equation}
    \frac{\frac{||\delta {\bf x}||}{||{\bf x} + \delta {\bf x}||}}{\frac{||\delta \mathrm{A}||}{||\mathrm{A}||}} \leq \kappa(\mathrm{A}) .
\end{equation}

\vspace{0.5cm}
\noindent
3) When both ${\bf b}$ and $\mathrm{A}$ have errors, $\delta {\bf b}$ and $\delta \mathrm{A}$, one can also derive the inequality when both ${\bf b}$ and $\mathrm{A}$ have errors, $\delta {\bf b}$ and $\delta \mathrm{A}$:
\begin{equation}
    \frac{\frac{||\delta {\bf x}||}{||{\bf x}||}}{\frac{||\delta \mathrm{A}||}{||\mathrm{A}||} + \frac{||\delta {\bf b}||}{||{\bf b}||}} \leq \frac{\kappa(\mathrm{A})}{1 - ||\mathrm{A}^{-1} \delta \mathrm{A}||} .
\end{equation}
We assumed $||\mathrm{A}^{-1} \delta \mathrm{A}||<1$.

\vspace{1cm}
In summary, the condition number provides an upper bound for the relative error in the linear system.  
When the condition number $\kappa(\mathrm{A})$ is small, tiny perturbations $\delta \mathrm{A}$ and/or $\delta {\bf b}$ lead to only small relative errors in ${\bf x}$, and the linear system is stable.  
When the condition number is large, however, even tiny perturbations may induce huge errors, making the linear system unstable, even if $\mathrm{A}$ is invertible.  

\section{Ridge regression in the PCA basis}
\label{sec:appendix_pcr}

\subsection{Expression of the Pearson coefficient}

Let us consider the Ridge regression solution, Eq.~\eqref{eq:w_Ridge}, in the PCA setting.
Using Eq.~\eqref{eq:U_rho_YX} and $(\mathrm{C} + \alpha \mathrm{I})^{-1} = \mathrm{U} (\mathrm{\Lambda} + \alpha \mathrm{I})^{-1} \mathrm{U}^T$, we get
\begin{eqnarray}
    \hat {\bf Y} &=& \mathrm{X} \hat {\bf w}_{\rm Ridge} = \mathrm{X} (\mathrm{C} + \alpha \mathrm{I})^{-1} \Rb{{\bf X}}{Y} \nonumber \\
    &=& \mathrm{X'} (\Lambda + \alpha \mathrm{I})^{-1} \mathrm{\Sigma} \Rb{\tilde{\bf X}}{Y} \nonumber \\
    &=& \sum_{f=1}^M \R{\tilde{X}^{(f)}}{Y} \left( \frac{\lambda^{(f)}}{\lambda^{(f)} + \alpha} \right) \tilde {\bf X}^{(f)} .
    \label{eq:alpha_vs_lambda_full}
\end{eqnarray}

Using the PCA basis, we obtain the Pearson coefficient between the ground truth $Y_i$ and the prediction $\hat Y_i$
\begin{eqnarray}
    \R{Y}{\hat Y} &=& \frac{1}{N_\mathcal{S} \sqrt{\mathrm{Var}[\hat Y]}} \sum_{i \in \mathcal{S}} Y_i \hat Y_i = \frac{1}{N_\mathcal{S} \sqrt{\mathrm{Var}[\hat Y]}} \hat {\bf Y}^T {\bf Y} \nonumber \\
    &=& \frac{1}{\sqrt{\mathrm{Var}[\hat Y]}} \hat {\bf w}^T \Rb{{\bf X}}{Y} 
\end{eqnarray}
and 
\begin{equation}
        \mathrm{Var}[\hat Y] =  \frac{1}{N_\mathcal{S}} \hat {\bf Y}^T \hat {\bf Y} = \hat {\bf w}^T \mathrm{C} \hat {\bf w} ,
\end{equation}
where we used Eqs.~(\ref{eq:prediction}), (\ref{eq:correlation_matrix_XX}), and (\ref{eq:Pearson_XY}).

By plugging in the Ridge regression weights, $\hat{\bf w}_{\rm Ridge}$, from Eq.~(\ref{eq:w_Ridge}), we get
\begin{equation}
    \R{Y}{\hat Y} = \frac{\Rb{{\bf X}}{Y}^T (\mathrm{C}+\alpha \mathrm{I})^{-1} \Rb{{\bf X}}{Y}}{\sqrt{\Rb{{\bf X}}{Y}^T (\mathrm{C}+\alpha \mathrm{I})^{-1} \mathrm{C} (\mathrm{C}+\alpha \mathrm{I})^{-1} \Rb{{\bf X}}{Y}}} .
\end{equation}

Using the PCA basis, namely Eqs.~(\ref{eq:PCA_diagonal}) and (\ref{eq:U_rho_YX}), we finally arrive at Eq.~\eqref{eq:rho_YY_Ridge_PCA} in the main text.

\subsection{Role of the regularization parameter}

Looking at the final expression in Eq.~\eqref{eq:alpha_vs_lambda_full}, we see that, or a given $\alpha > 0$, when $\lambda^{(f)} \gg \alpha$, the factor involving the eigenvalues becomes $\lambda^{(f)}/(\lambda^{(f)} + \alpha) \simeq 1$. On the other hand, when $\lambda^{(f)} \ll \alpha$, $\lambda^{(f)}/(\lambda^{(f)} + \alpha) \simeq 0$, hence such features are suppressed in the sum.
Therefore, through the factor $\lambda^{(f)}/(\lambda^{(f)} + \alpha)$, $\alpha$ acts as a soft threshold, determining whether we include a feature or not.  

It is then easy to find the Pearson coefficient 
\begin{equation}
    \R{Y}{\hat Y} = \frac{\sum_{f=1}^M (\R{\tilde X^{(f)}}{Y})^2\left(\frac{\lambda^{(f)}}{\lambda^{(f)}+\alpha}\right)}{\sqrt{\sum_{f=1}^M (\R{\tilde X^{(f)}}{Y})^2\left(\frac{\lambda^{(f)}}{\lambda^{(f)}+\alpha}\right)^2}} .
    \label{eq:rho_YY_Ridge_PCA}
\end{equation}
We observe that the same threshold factor appears again in the expression, offering a clear interpretation of the role of Ridge regularization: in the PCA basis, the regularization term softly suppresses modes associated with eigenvalues smaller than $\alpha$.
Thus, smaller eigenvalues, potentially associated to multicollinearity, are systematically and softly suppressed as $\alpha$ increases.

It turns out, however, that the Ridge regularization does not provide a practical benefit  within the PCR scheme.
We found, in fact, that the order of the features is unaltered for small $P$, both when sorting against $\lambda$ and against $R$.
Hence, $\alpha$ does not affect feature selection discussed in Sec.~\ref{sec:pcr}.
Moreover, the maximum correlation achieved when all the $M$ features are included decreases with increasing $\alpha$.

\subsection{Suppression of the oscillatory behavior of the weights}
\label{sec:oscillation_Ridge_PCA}

By considering the Ridge regression problem in the PCA basis, we can now provide a simple theoretical argument to illustrate how regularization suppresses the oscillatory behavior of the weights.

We start from Eq.~(\ref{eq:w_Ridge}) and use the eigenvalue decomposition of $(\mathrm{C} + \alpha \mathrm{I})^{-1}$, which allows us to express $\hat{\bf w}_{\rm Ridge}$ as a sum of projections of the Pearson coefficient vector $\Rb{X}{Y}$ onto each eigenmode:
\begin{equation}
    \hat{\bf w}_{\rm Ridge} = \sum_{f=1}^M (\lambda^{(f)}+\alpha)^{-1}{\bf u}^{(f)} {{\bf u}^{(f)}}^T \Rb{X}{Y} .
\end{equation}
As discussed in Sec.~\ref{sec:condition_number}, the oscillatory behavior can be highlighted in the two-feature model, namely,
\begin{equation}
    \hat{\bf w}_{\rm Ridge} = \left[ \frac{{\bf u}^{(1)} {{\bf u}^{(1)}}^T}{\lambda_{\rm max}+\alpha} + \frac{{\bf u}^{(2)} {{\bf u}^{(2)}}^T}{\lambda_{\rm min}+\alpha}  \right] \Rb{X}{Y} ,
\end{equation}
where $\Rb{X}{Y} = [R^{(1)}, R^{(2)}]^T$ and $\hat{\bf w}_{\rm Ridge} = [\hat w_{\rm Ridge}^{(1)}, \hat w_{\rm Ridge}^{(2)}]^T$.
This is further rewritten as 
\begin{eqnarray}
    \hat w_{\rm Ridge}^{(1)} &=& \frac{R^{(1)}+R^{(2)}}{2(\lambda_{\rm max}+\alpha)} +  \frac{R^{(1)}-R^{(2)}}{2(\lambda_{\rm min}+\alpha)} ,\\
    \hat w_{\rm Ridge}^{(2)} &=& \frac{R^{(1)}+R^{(2)}}{2(\lambda_{\rm max}+\alpha)} -  \frac{R^{(1)}-R^{(2)}}{2(\lambda_{\rm min}+\alpha)} .
\end{eqnarray}
When $\alpha \to 0$, $\hat{\mathbf{w}}_{\rm Ridge} \to \hat{\mathbf{w}}_{\rm OLS}$, and one observes the oscillatory behavior, $\hat{w}_{\rm OLS}^{(1)} \simeq -\hat{w}_{\rm OLS}^{(2)}$, with large magnitude, arising from the mode associated with $\lambda_{\rm min} \to 0$.
For sufficiently large $\alpha$, the contribution from such mode is suppressed, thereby mitigating the oscillatory behavior.


%

\end{document}